\documentclass[12pt]{SelfArx} 
\usepackage[english]{babel} 
\usepackage{amsfonts,amssymb,amsbsy,amsmath,amsthm}
\usepackage{listings}
\usepackage{lscape}
\usepackage{multirow}
\usepackage{colortbl,hhline}
\usepackage{here}
\usepackage{url}
\usepackage[numbers]{natbib}

\theoremstyle{definition}

\theoremstyle{plain}

\newcommand{\g}[1]{\boldsymbol{\mathrm{#1}}}
\newcommand{\tp}{\mathrm{t}}
\newcommand{\diag}{\mathrm{diag}}
\newcommand{\idif}{\,\mathrm{d}}
\newcommand{\SE}{\mathrm{SE}}
\newcommand{\rank}{\mathrm{rank}}

\definecolor{color1}{RGB}{0,0,90} 
\definecolor{color2}{RGB}{0,20,20} 

\usepackage{hyperref} 
\hypersetup{hidelinks,colorlinks,breaklinks=true,urlcolor=red,citecolor=color1,linkcolor=color1,bookmarksopen=false,
	pdftitle={A modified maximum contrast method for unequal sample sizes in pharmacogenomic studies},
	pdfauthor={Kengo Nagashima, Yasunori Sato, Chikuma Hamada}}

\JournalInfo{\emph{Stat Appl Genet Mol Biol} 2011; \textbf{10}(1): Article 41.} 
\Archive{DOI: 10.2202/1544-6115.1560} 

\PaperTitle{A modified maximum contrast method for unequal sample sizes in pharmacogenomic studies}
\ShortTitle{Modified maximum contrast method for unequal sample sizes}

\Authors{Kengo Nagashima\textsuperscript{1}*, Yasunori Sato\textsuperscript{2}, and Chikuma Hamada\textsuperscript{3}} 
\affiliation{\footnotesize{\textsuperscript{1}\textit{Faculty of Pharmaceutical Sciences, Josai University, Japan.}}}
\affiliation{\footnotesize{\textsuperscript{2}\textit{Department of Biostatistics, Harvard School of Public Health, USA.}}}
\affiliation{\footnotesize{\textsuperscript{3}\textit{Faculty of Engineering, Tokyo University of Science, Japan.}}}
\affiliation{\footnotesize{*\textbf{Corresponding author}: nshi1201@gmail.com}}
\affiliation{}
\affiliation{This is a preprint/accepted version (30-Jul-2011) of the following manuscript: Nagashima K, Sato Y, Hamada C. A modified maximum contrast method for unequal sample sizes in pharmacogenomic studies. \emph{Statistical Applications in Genetics and Molecular Biology} 2011; \textbf{10}(1): Article 41. DOI: \href{https://doi.org/10.2202/1544-6115.1560}{10.2202/1544-6115.1560}.}

\newcommand{\keywordname}{Keywords} 

\Abstract{
\footnotesize{
In pharmacogenomic studies, biomedical researchers commonly analyze the association between genotype and biological response by using the Kru\-skal--Wallis test or one-way analysis of variance (ANOVA) after logarithmic transformation of the obtained data.
However, because these methods detect unexpected biological response patterns, the power for detecting the expected pattern is reduced.
Previously, we proposed a combination of the maximum contrast method and the permuted modified maximum contrast method for unequal sample sizes in pharmacogenomic studies.
However, we noted that the distribution of the permuted modified maximum contrast statistic depends on a nuisance parameter $\sigma^2$, which is the population variance.
In this paper, we propose a modified maximum contrast method with a statistic that does not depend on the nuisance parameter.
Furthermore, we compare the performance of these methods via simulation studies.
The simulation results showed that the modified maximum contrast method gave the lowest false-positive rate; therefore, this method is powerful for detecting the true response patterns in some conditions.
Further, it is faster and more accurate than the permuted modified maximum contrast method.
On the basis of these results, we suggest a rule of thumb to select the appropriate method in a given situation.
}
}

\begin{document}

\flushbottom 

\maketitle 


\thispagestyle{empty} 


\section{Introduction}
Interindividual variations in drug efficacy and side effects pose a serious problem in medicine.
These variations are influenced by factors such as drug-metabolizing enzymes, drug transporters, and drug targets (e.g., receptors).
For many medications, these factors develop partly due to genetic polymorphisms \citep{Evans2001,Evans2003}.
In fact, genomic biomarkers are sometimes used to modify drug responses and reduce side effects by controlling the medication or dose according to the genotype \citep{UGT1A1,Wilkinson}, but more of these biomarkers need to be identified.

It is difficult to identify genomic biomarkers according to whether patients will respond positively, be non-responders, or experience adverse reactions to the same medication and dose.
Therefore, many pharmacogenomic studies have been launched worldwide, such as a pharmacokinetic (PK) study including analyses of single-nucleotide polymorphisms (SNPs) in a candidate gene or a genome-wide approach.
With the completion of the International HapMap Project \citep{HAPMAP} and the availability of powerful array-based SNP-typing platforms, the genome-wide approach has become the popular strategy for identifying susceptibility to and drug-response genes in common diseases.

To identify SNPs related to drug metabolism, biomedical researchers usually test the null hypothesis ($H_0$) that there is no difference between the genotype in the location parameters of the distribution of PK parameters such as area under the blood concentration--time curve ($AUC$), maximum drug concentration ($C_{\max}$), and half-life period ($t_{1/2}$).
Researchers commonly use the Kruskal--Wallis test \citep{KW} or one-way analysis of variance (ANOVA) after logarithmic transformation of the obtained data.
On the basis of the statistical significance of these tests, they then visually check the response patterns between the PK parameters and genotypes for the expected biological response patterns.

\begin{figure}[ht]
\begin{center}
\includegraphics[clip,scale=.2]{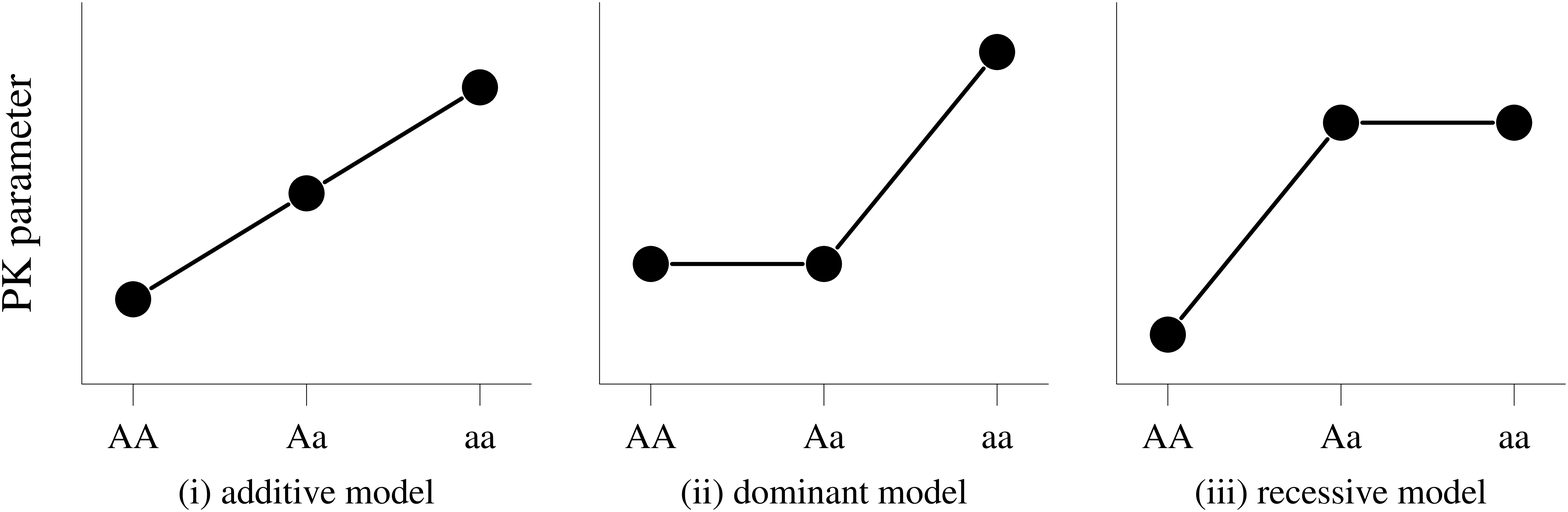}\\
\includegraphics[clip,scale=.2,trim=0 0 1150 0]{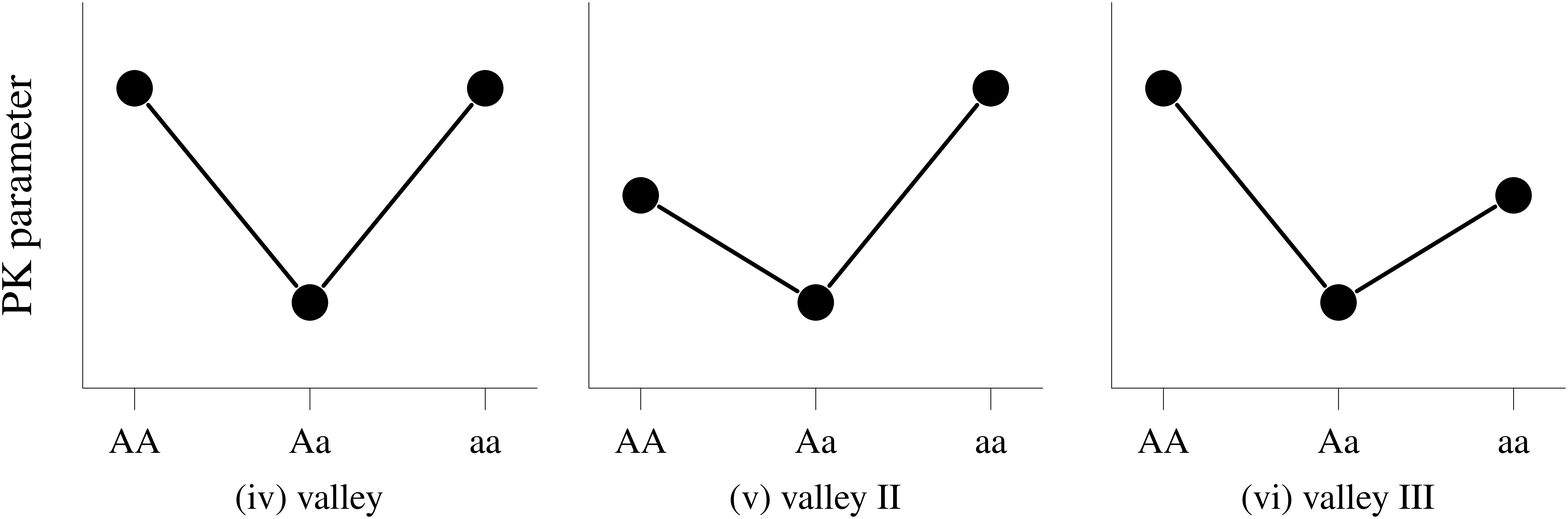}
\end{center}
\caption{Response patterns between a PK parameter and genotype}
\label{fig:rpattern}
\end{figure}

For additive, dominant, and recessive patterns, the PK parameters monotonically increase or decrease in the wider sense with the number of alleles, as shown in Figure \ref{fig:rpattern}, i--iii.
However, because there are two degrees of freedom in the Kruskal--Wallis test, unexpected non-monotonic biological response patterns can be detected (see Figure \ref{fig:rpattern}, iv).
Thus, this commonly used approach has disadvantages when screening for the true PK-related genes.

In a previous study, we proposed contrast statistic-based methods \citep{MMCM} for screening PK-related genes in genome-wide studies.
We applied the maximum contrast method \citep{MCM,Wakana} but found that this method is inferior for detecting specific response patterns in unequal sample sizes.
In pharmacogenomic studies, the sample size of each genotypic group was rather different.
Thus, we proposed the permuted modified maximum contrast method for unequal sample sizes and combined it with the maximum contrast method.
These methods can consider a specific alternative hypothesis for monotonic response patterns (Figure \ref{fig:rpattern}, i--iii; see Section 2).

However, we noted that the distribution of the permuted modified maximum contrast statistic under the overall null hypothesis depends on a nuisance parameter $\sigma^2$, which is the population variance.
Therefore, in this paper, we propose a modified maximum contrast method with a statistic that does not depend on this parameter (see Section 3).

Further, using simulation studies, we compare the performance of the Kru\-skal--Wallis test, the maximum contrast method, and the modified maximum contrast method in pharmacogenomic studies (see Section 4).
Finally, we compare the computational speed and accuracy of the permuted modified maximum contrast method and the modified maximum contrast method (see Section 5).

\section{Contrast statistic-based methods}

\subsection{Notation and assumptions}
Herein, we consider the typical one-way fixed analysis of variance model with unequal sample sizes. 
The random variable $X_{ij}$ indicates the observed response (PK parameter) of the $j$-th subject in the $i$-th group (genotype), and $Y_{ij}$ is the logarithmic transformation of the observed response.
We assume that
\begin{equation}
\log X_{ij}=Y_{ij} \overset{\mathrm{i.i.d.}}{\sim} N(\mu_i, \sigma^2),~~
i = 1, 2, \ldots, a,~~
j = 1, 2, \ldots, n_i
,
\label{equ:assumption}
\end{equation}
where $\mu_i$ is the population mean of the $i$-th group, and $\sigma^2$ is the population variance.
Under the assumption shown in Equation \ref{equ:assumption}, the sample mean vector of each group, $\bar{\g{Y}} = (\bar{Y}_1, \bar{Y}_2, \ldots, \bar{Y}_i, \ldots, \bar{Y}_a)^{\tp}$, follows the $a$-variate normal distribution $N_a(\g{\mu}, \sigma^2\g{D})$, where $\g{\mu} = (\mu_1, \mu_2, \ldots, \mu_i, \ldots, \mu_a)^{\tp}$ and $\g{D} = \diag(1/n_1, 1/n_2, \ldots, 1/n_i, \ldots, 1/n_a)$.
Of note, $\diag(\,)$ indicates a diagonal matrix with diagonal elements in parentheses and superscript ``t'' indicates the transpose of a matrix.

There are most commonly three genotypes considered for the relationship between SNPs and the PK parameters in pharmacogenomic studies: $i$ = 1 (AA), 2 (Aa), and 3 (aa), where ``A'' and ``a'' are the major and minor alleles, respectively.
Moreover, although the exact distributions of the PK parameters are often unknown, they are empirically modeled using the assumption of a log-normal distribution, because the PK parameters must not be negative, and the normal distribution does not satisfy this condition; in addition, the distribution of the estimated PK parameters is often right-skewed, which is compatible with a log-normal distribution \citep{PKPD}.

\subsection{The maximum contrast method}
The maximum contrast method for dose-response studies has been previously discussed by \citet{MCM} and \citet{Wakana}.
Both of these groups considered the maximum contrast statistic, $T_{\max}$, for testing the overall null hypothesis, $H_0$, versus the ordered or monotonic multiple alternative hypotheses, $H_1$.
\begin{equation}
\begin{cases}
H_0 : & \mu_1=\mu_2=\ldots=\mu_i=\ldots=\mu_a\\
H_1 : & \g{C}\g{\mu}>\g{0}
\end{cases}
\label{equ:hypotheses}
\end{equation}
To specify alternative hypotheses, it is necessary to define the constants as $\g{C}=(\g{c}_1,\g{c}_2,\ldots,\g{c}_k,\ldots,\g{c}_m)^{\tp}$, where $\g{c}_k=(c_{k1},c_{k2},\ldots,c_{ki},\ldots,c_{ka})^{\tp}$ subject to $\sum_{i=1}^{a}c_{ki}=0$ and $m$ is the number of alternative hypotheses.
The matrix $\g{C}$ is referred to as the contrast coefficient matrix, and the element $\g{c}_k$ is referred to as the $k$-th contrast coefficient vector.

In a typical pharmacogenomic study, the association between the PK parameters and genotypes is modeled according to the response patterns in Figure \ref{fig:rpattern}, i--iii, and the maximum contrast method is subsequently applied to the three contrast statistics with the following contrast coefficient matrix
\begin{equation}
\g{C}=
\begin{pmatrix}
\g{c}_1 & \g{c}_2 & \g{c}_3
\end{pmatrix}^{\tp}
=
\begin{pmatrix}
-1/2 & -1/3 & -2/3 \\
   0 & -1/3 &  1/3 \\
 1/2 &  2/3 &  1/3 \\
\end{pmatrix}^{\tp}.
\label{equ:pgcontrast}
\end{equation}
The first contrast coefficient vector corresponds to an additive model, the second to a recessive model, and the third to a dominant model.
In terms of the matrix $\g{C}$, Equation \ref{equ:pgcontrast} implies that the alternative hypotheses are $H_1: \mu_1 < \mu_2 < \mu_3$, $\mu_1 = \mu_2 < \mu_3$, and $\mu_1 < \mu_2 = \mu_3$.

The maximum contrast statistic is defined as
\begin{equation}
T_{\max} = \max_{k=1,2,\ldots,m}\{T_k\},~~
T_k=
\frac{Z_k}{\sqrt{\left(\gamma\frac{V}{\sigma^2}\right)\Big/\gamma}}=
\frac{\g{c}_k^{\tp}\bar{\g{Y}}}{\sqrt{V\g{c}_k^{\tp}\g{D}\g{c}_k}}
~,
\label{equ:tmax}
\end{equation}
where the random variable $Z_k = \g{c}_k^{\tp}\bar{\g{Y}}\Big/\sqrt{\sigma^2\g{c}_k^{\tp}\g{D}\g{c}_k}$ follows the $m$-variate normal distribution\\ $N_m(\g{c}_k^{\tp}\g{\mu}\Big/\sqrt{\sigma^2\g{c}_k^{\tp}\g{D}\g{c}_k}, 1^2)$, $V = \frac{1}{\gamma} \sum_{i=1}^a \sum_{j=1}^{n_i}(Y_{ij}-\bar{Y}_i)^2$ is an unbiased estimator of $\sigma^2$, and $\gamma = \sum_{i=1}^a (n_i - 1)$ is the degrees of freedom for $V$.

The simultaneous distribution of the random vector $\g{T} = (T_1, T_2, \ldots, T_k, \ldots, T_m)^{\tp}$ is the non-central $m$-variate $t$-distribution $t_m(\g{\Sigma}_T, \gamma, \g{\lambda}_T)$, where
\begin{equation*}
\g{\lambda}_T=\left\{\g{c}_k^{\tp}\g{\mu}\Big/\sqrt{\sigma^2\g{c}_k^{\tp}\g{D}\g{c}_k}\right\}_{1\leq k\leq m}
\end{equation*}
is a non-central parameter vector, and $\g{\Sigma}_T$ is a positive semi-definite covariance matrix represented as
\begin{equation*}
\g{\Sigma}_T=
\left\{
\frac{\g{c}_k^{\tp}\g{D}\g{c}_l}{\sqrt{\g{c}_k^{\tp}\g{D}\g{c}_k}\sqrt{\g{c}_l^{\tp}\g{D}\g{c}_l}}
\right\}_{1\leq k,l \leq m}
~.
\end{equation*}
Note that $\g{\Sigma}_T^{-1}$ does not exist if $|\g{\Sigma}_T|=0$ for a given covariance matrix.
For such singular multivariate $t$-distribution distributions, the probability mass is concentrated on a linear subspace.
Fortunately, the integration method for these distributions has been proposed by Genz and Bretz \cite{GenzBretz2009}, which separates the linear subspace and transforms the integration region by using separation-of-variables transformations.

The $P$-value of the maximum contrast method can be derived as follows \citep{GenzBretz1999}:
\begin{equation}
\begin{split}
P\mbox{-value} &=
\Pr(T_{\max} \geq t_{\max} \mid H_0)=
1 - \Pr(T_{\max} < t_{\max} \mid H_0) \\&=
1 - \Pr(T_1 < t_{\max}, T_2 < t_{\max}, \ldots, T_k < t_{\max}, \ldots, T_m < t_{\max} \mid H_0)
~,
\end{split}
\label{equ:pvalue}
\end{equation}
where $t_{\max}$ is the observed value of the test statistic.
To calculate Equation \ref{equ:pvalue}, we must integrate the simultaneous distribution of $\g{T}$ under the overall null hypothesis.
Therefore, Equation \ref{equ:pvalue} is given by the integral
\begin{equation}
P\mbox{-value} =
1-\int_{-\infty}^{t_{\max}}\int_{-\infty}^{t_{\max}}\cdots\int_{-\infty}^{t_{\max}}
t_m(\g{T}=\g{t} \mid \g{\Sigma}_T, \gamma)
\idif \g{t}.
\label{equ:integratepval}
\end{equation}
In this article, Equation \ref{equ:integratepval} is calculated using the randomized quasi-Monte Carlo method for integration \citep{GenzBretz1999,GenzBretz2002,GenzBretz2009}.
In addition, the coefficient vector for the contrast statistic with the maximum value $\g{c}_{t_{\max}}=\{\g{c}_k \mid t_k=t_{\max}\}$ is then selected as the true response pattern that best fits the observed data.

Designs with equal sample sizes are often used in dose-response studies.
In contrast, in pharmacogenomic studies the sample size of each group is not controlled and the population is in Hardy--Weinberg equilibrium.
Therefore, these studies are likely to have unequal sample sizes for different genotypes, and a minor allele frequency (MAF) of less than 0.5, and most commonly around 0.2.

\begin{figure}[htb]
\begin{center}
\includegraphics[clip,scale=.3]{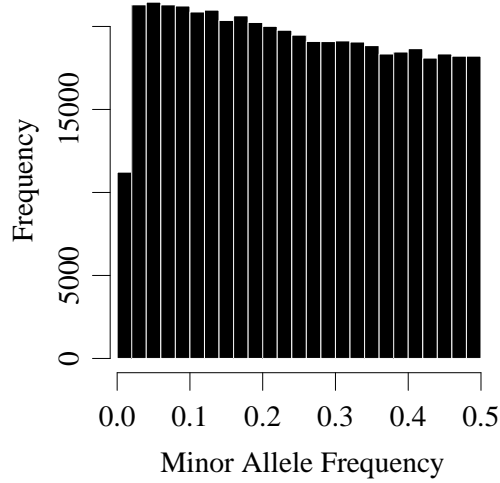}
\end{center}
\caption{Distribution of MAF from JSNP public database (Hirakawa et al. \cite{JSNP})}
\label{fig:maf}
\end{figure}

In cases with unequal sample sizes, the denominator of the contrast statistic from Equation \ref{equ:tmax},
\[
\sqrt{V\left(\frac{c_{k1}^2}{n_1}+\frac{c_{k2}^2}{n_2}+\frac{c_{k3}^2}{n_3}\right)},
\]
is overestimated at specific contrast coefficient vectors, although the statistic of this variance estimate is robust.
Thus, using only the maximum contrast method is insufficient for detecting the true response pattern in pharmacogenomic studies.

\subsection{The permuted modified maximum contrast method}
Since the maximum contrast method should not be used alone in pharmacogenomic studies, it has been proposed that the permuted modified maximum contrast method should instead be used for such cases with unequal sample sizes \citep{MMCM}, with statistic
\begin{equation*}
M_{\max} = \max_{k=1,2,\ldots,m}\{M_k\},~~
M_k = \frac{\g{c}_k^{\tp}\bar{\g{Y}}}{\sqrt{\g{c}_k^{\tp}\g{c}_k}}.
\end{equation*}
This statistic can be used to test the hypotheses in Equation \ref{equ:hypotheses}, and the $P$-value can be defined similarly to that in Equation \ref{equ:pvalue}, $\Pr(M_{\max} \geq m_{\max} \mid H_0)$.
Moreover, this method selects the coefficient vector that best fits the observed data by $\g{c}_{m_{\max}}=\{\g{c}_k \mid m_k=m_{\max}\}$.
The simultaneous distribution of the random vector $\g{M} = (M_1, M_2, \ldots, M_k, \ldots, M_m)^{\tp}$ is the $m$-variate normal distribution $N_m(\g{\lambda}_M, \g{\Sigma}_M)$, where $\g{\lambda}_M=\left\{\g{c}_k^{\tp}\g{\mu}\Big/\sqrt{\g{c}_k^{\tp}\g{c}_k}\right\}_{1\leq k\leq m}$ is a population mean vector, and $\g{\Sigma}_M$ is a positive semi-definite covariance matrix represented as
\begin{equation}
\g{\Sigma}_M=
\left\{
\frac{\sigma^2\g{c}_k^{\tp}\g{D}\g{c}_l}{\sqrt{\g{c}_k^{\tp}\g{c}_k}\sqrt{\g{c}_l^{\tp}\g{c}_l}}
\right\}_{1\leq k,l \leq m}.
\label{equ:mcovariance}
\end{equation}

It follows from Equation \ref{equ:mcovariance} that the statistic $M_{\max}$ depends on the value of the nuisance parameter $\sigma^2$ under the overall null hypothesis, such that the exact distribution is unknown.
In a previous study, an approximate $P$-value was calculated by the permutation method \citep{WestfallYoung} following Algorithm 1.\\

\noindent \textbf{Algorithm 1}: Permutation method for the statistic $M_{\max}$.
\begin{enumerate}[leftmargin=*]
\item Initialize counting variable: $COUNT = 0$.\\
Input parameters: $NRESAMPMIN$ (minimum resampling count, set to 1000), $NRESAMPMAX$ (maximum resampling count), and $\epsilon$ (absolute error tolerance).
\item Calculate $m_{\max}$, the observed value of the test statistic.
\item Let $y_{ij}^{(r)}$ denote the data, which are sampled without replacement and independently from observed value $y_{ij}$.
Here, $r$ is the resampling index $(r = 1, 2, \ldots, NRESAMP)$.
\item Calculate $m_{\max}^{(r)}$ from $y_{ij}^{(r)}$.
If $m_{\max}^{(r)} > m_{\max}$, then increment the counting variable: $COUNT = COUNT + 1$.
Calculate the approximate $P$-value, $\hat{p}^{(r)} = COUNT / r$, and the simulation standard error, $\hat{\sigma}^{(r)}=\SE(\hat{p}^{(r)})= \sqrt{\hat{p}^{(r)}(1-\hat{p}^{(r)})/r}$.
\item Repeat steps 3 and 4 if $r > NRESAMPMIN$ and $3.5 \hat{\sigma}^{(r)} < \epsilon$ (corresponding to an approximate confidence level of 99.95\%; this is the accuracy of the randomized quasi-Monte-Carlo method of \citet{GenzBretz2002}) or $NRESAMPMAX$ times.
Output the approximate $P$-value, $\hat{p}^{(r)}$, and the standard error, $\SE(\hat{p}^{(r)})$.
\end{enumerate}

\section{The proposed method}
\subsection{The modified maximum contrast method}
In this section, we propose a modified maximum contrast statistic
\begin{equation}
S_{\max} = \max_{k=1,2,\ldots,m}\{S_k\},~~
S_k=
\frac{Z_k^{\prime}}{\sqrt{\left(\gamma\frac{V}{\sigma^2}\right)\Big/\gamma}}=
\frac{\g{c}_k^{\tp}\bar{\g{Y}}}{\sqrt{V\g{c}_k^{\tp}\g{c}_k}},
\label{equ:smax}
\end{equation}
where $Z_k^{\prime} = \g{c}_k^{\tp}\bar{\g{Y}}\Big/\sqrt{\sigma^2\g{c}_k^{\tp}\g{c}_k}$ is the random variable $N_m(\g{c}_k^{\tp}\g{\mu}\Big/\sqrt{\sigma^2\g{c}_k^{\tp}\g{c}_k},\g{c}_k^{\tp}\g{D}\g{c}_k\Big/\g{c}_k^{\tp}\g{c}_k)$.
Since the distribution of the statistic in Equation \ref{equ:smax} is not dependent on $\sigma^2$ under the overall null hypothesis, the $P$-value is calculated using the randomized quasi-Monte Carlo method for integration, which improves the computational speed and accuracy (see Section 5).
The modified maximum contrast statistic can be used to test the hypotheses in Equation \ref{equ:hypotheses}, and the $P$-value is defined similarly to that in Equation \ref{equ:pvalue}, $\Pr(S_{\max} \geq s_{\max} \mid H_0)$.
Moreover, this method can select the coefficient vector that best fits the observed data by $\g{c}_{s_{\max}}=\{\g{c}_k \mid s_k=s_{\max}\}$.
The simultaneous distribution of the random vector $\g{S} = (S_1, S_2, \ldots, S_k, \ldots, S_m)^{\tp}$ is the non-central $m$-variate $t$-distribution $t_m(\g{\Sigma}_S, \gamma, \g{\lambda}_S)$, where $\g{\lambda}_S=\left\{\g{c}_k^{\tp}\g{\mu}\Big/\sqrt{\sigma^2\g{c}_k^{\tp}\g{c}_k}\right\}_{1\leq k\leq m}$ is a population mean vector, and $\g{\Sigma}_S$ is a positive semi-definite covariance matrix represented as
\begin{equation*}
\g{\Sigma}_S=
\left\{
\frac{\g{c}_k^{\tp}\g{D}\g{c}_l}{\sqrt{\g{c}_k^{\tp}\g{c}_k}\sqrt{\g{c}_l^{\tp}\g{c}_l}}
\right\}_{1\leq k,l \leq m}
.
\end{equation*}

Therefore, $\g{S}$ follows the $m$-variate $t$-distribution $t_m(\g{\Sigma}_S, \gamma)$ under the overall null hypothesis.

\subsection{Difference between the maximum contrast method and the modified maximum contrast method}
In this section, we illustrate an important difference between the maximum contrast method and the modified maximum contrast method.
In particular, the difference between the statistics $T_{\max}$ and $S_{\max}$ is that they are respectively with and without the matrix $\g{D} = \diag(1/n_1, 1/n_2, \ldots, 1/n_i, \ldots, 1/n_a)$ in the dominator of Equations \ref{equ:tmax} and \ref{equ:smax}.
Of note, this difference affects the properties of both methods.

Let the non-central $m$-variate $t$ integral be given by
\begin{equation*}
T_m(\g{a},\g{b}; \g{\Sigma},\gamma,\g{\lambda})=
\int_{a_1}^{b_1}\int_{a_2}^{b_2}\cdots\int_{a_m}^{b_m}
t_m(\g{X}=\g{x} \mid \g{\Sigma},\gamma,\g{\lambda})
\idif\g{x},
\end{equation*}
where $[\g{a}, \g{b}],-\infty \leq a_k < b_k \leq \infty, k=1,2,\ldots,m$.
The critical values that correspond to a significance level $\alpha$ can then be defined by
\begin{equation*}
\begin{split}
u_{\alpha}&=\{u \mid 1-T_m(-\g{\infty},\g{u}; \g{\Sigma}_T,\gamma,\g{0})=\alpha\},\\
v_{\alpha}&=\{v \mid 1-T_m(-\g{\infty},\g{K}^{-1}_S\g{v}; \g{\Sigma}_T,\gamma,\g{0})=\alpha\},
\end{split}
\end{equation*}
because the cumulative distribution function of the modified maximum contrast statistic can be written as
\begin{equation*}
\begin{split}
\Pr(S_{\max} \leq v \mid H_1)
&=
T_m(-\g{\infty},\g{v}; \g{\Sigma}_S,\gamma,\g{\lambda}_S)\\
&=
T_m(-\g{\infty},\g{K}^{-1}_S\g{v}; \g{\Sigma}_T,\gamma,\g{\lambda}_T),
\end{split}
\end{equation*}
where
\begin{equation*}
\g{K}^{-1}_S=\diag\left\{\sqrt{\frac{\g{c}_k^{\tp}\g{c}_k}{\g{c}_k^{\tp}\g{D}\g{c}_k}}\right\}_{1\leq k \leq m},
\end{equation*}
$-\g{\infty}=(-\infty,-\infty,\ldots,-\infty)^{\tp}$, $\g{u}=(u,u,\ldots,u)^{\tp}$, $\g{v}=(v,v,\ldots,v)^{\tp}$, and $\g{0}=(0,0,\ldots,0)^{\tp}$.
We now have a power function of the maximum contrast statistic that is defined by
\begin{equation}
\beta_T(\g{\mu};\g{C},\g{D})=1-T_m(-\g{\infty},\g{u}_{\alpha}; \g{\Sigma}_T,\gamma,\g{\lambda}_T)
,
\label{equ:powert}
\end{equation}
and a power function of the modified maximum contrast statistic that is defined by
\begin{equation}
\beta_S(\g{\mu};\g{C},\g{D})=1-T_m(-\g{\infty},\g{K}^{-1}_S\g{v}_{\alpha}; \g{\Sigma}_T,\gamma,\g{\lambda}_T)
,
\label{equ:powers}
\end{equation}
where $\g{u}_{\alpha}=(u_{\alpha},u_{\alpha},\ldots,u_{\alpha})^{\tp}, \g{v}_{\alpha}=(v_{\alpha},v_{\alpha},\ldots,v_{\alpha})^{\tp}$.
The critical values $\g{u}_{\alpha}$ are symmetric, whereas the critical values $\g{K}^{-1}_S\g{v}_{\alpha}$ are asymmetric in Equations \ref{equ:powert} and \ref{equ:powers}.
Therefore, the difference between the statistics $T_{\max}$ and $S_{\max}$ is that the rejection region is respectively equivalent to or not equivalent to each contrast statistic.
In other words, the statistic $S_{\max}$ gives priority to a contrast statistic $S_k$ that satisfies the equation below:
\begin{equation}
\min_k \left\{\sqrt{\frac{\g{c}_k^{\tp}\g{c}_k}{\g{c}_k^{\tp}\g{D}\g{c}_k}}\right\}.
\label{equ:smaxpriority}
\end{equation}

Similarly, we define
\begin{equation}
\begin{split}
R_{\mathrm{TP}(T)}&=\Pr(T_{\max} \geq u_{\alpha}, \g{T} \leq T_{\mathrm{true}} \mid H_1),\\
R_{\mathrm{TP}(S)}&=\Pr(S_{\max} \geq v_{\alpha}, \g{S} \leq S_{\mathrm{true}} \mid H_1),
\end{split}
\label{equ:RTPTS}
\end{equation}
which is the probability for detecting the true response pattern $\g{\mu}$ among the
detected PK-related SNPs (positive predictive value), where the statistics $T_{\mathrm{true}}=\g{c}_{\mathrm{true}}^{\tp}\bar{\g{Y}}/\sqrt{V\g{c}_{\mathrm{true}}^{\tp}\g{D}\g{c}_{\mathrm{true}}}$ and $S_{\mathrm{true}}=\g{c}_{\mathrm{true}}^{\tp}\bar{\g{Y}}/\sqrt{V\g{c}_{\mathrm{true}}^{\tp}\g{c}_{\mathrm{true}}}$ satisfy $\g{c}_{\mathrm{true}}=a\g{\mu}$ with a finite constant $a\not=0$. However, it is important to note that evaluations of general cases are difficult.
We show some numerical examples of this in \ref{appendix:examplepower}.

\section{Simulation studies}
Here, we present the results of simulation studies to compare the methods.
We assessed the type I error rate, power, positive predictive value, and false-positive rate of the Kruskal--Wallis test, maximum contrast method, and modified maximum contrast method.

\subsection{Simulation conditions}
\label{sec:simcond}
The simulation conditions were almost identical to those used in the previous study (Sato et al., 2009).
The scenarios are similar to those of actual pharmacogenomic studies.
In particular, we were interested in the performance of all the three tests when the MAF decreases.
\begin{itemize}
\item
The MAF was set to 0.12, 0.25, 0.33, or 0.50.
It was uniformly distributed in [0.05, 0.5], according to the actual data (Hirakawa et al., \cite{JSNP}). The total sample size ($n$) was set to 100 or 300.
We assumed that the population was in Hardy--Weinberg equilibrium and set the sample size for each group as given in Table \ref{tab:MAFandN}.
\item
The genotypic response patterns examined were (i) additive, $\g{c}_1 = (-1/2,\, 0, 1/2)^{\tp}$; (ii) dominant, $\g{c}_2 = (-1/3, -1/3, 2/3)^{\tp}$; (iii) recessive, $\g{c}_3 = (-2/3, 1/3, 1/3)^{\tp}$ (expected response patterns); and (iv) valley, $\g{c}_4 = (1/3,\, -2/3,\, 1/3)^{\tp}$ (unexpected response pattern).\\
Under these conditions, we generated pseudo response values for each genotype by using random numbers from a normal distribution with mean $\mu_i$; that is, $N(\mu_i, 1^2)$, where $\mu_i = \Delta \times c_{ki}$, and $\Delta$ is a given coefficient for the effect sizes.
Here, $\Delta$ was set to 0.00, 0.25, 0.50, and 1.00.
\item
The maximum contrast and modified maximum contrast methods were applied with contrast coefficient matrix $\g{C}$ in Equation \ref{equ:pgcontrast}.
\item
The criteria to evaluate the performance of each method were $\hat{R}_{\mathrm{P}} = N_{\mathrm{P}} / N$, $\hat{R}_{\mathrm{TP}} = N_{\mathrm{TP}} / N$.
$R_{\mathrm{P}}$ is the probability to detect PK-related SNPs (power), whereas $R_{\mathrm{TP}}$ is the probability to detect the true response pattern among the detected PK-related SNPs (positive predictive value).
Here, $N$ is the repetition count of the simulation, $N_{\mathrm{P}}$ is the number of rejections by the hypothesis test, and $N_{\mathrm{TP}}$ is the number of detected true response patterns.
The two-tailed significance-level of each test was set to 0.05.
\item 
The Monte-Carlo simulations were repeated 20~000 times. This provided sufficient accuracy.
\end{itemize}

We performed the simulations in \texttt{R} and used the \texttt{R} function \texttt{pmvt()} to calculate the $P$-value for the maximum contrast and modified maximum contrast methods.

\begin{table}[htb]
\begin{center}
\caption{MAF and sample size for each group}
\label{tab:MAFandN}
\small
\begin{tabular}{rrrrrrrrr} \hline
\multicolumn{1}{c}{\multirow{2}{*}{MAF}} &  & \multicolumn{3}{c}{$n=100$} &  & \multicolumn{3}{c}{$n=300$} \\ \cline{3-5} \cline{7-9}
 &  & \multicolumn{1}{c}{$n_1$} & \multicolumn{1}{c}{$n_2$} & \multicolumn{1}{c}{$n_3$} &  & \multicolumn{1}{c}{$n_1$} & \multicolumn{1}{c}{$n_2$} & \multicolumn{1}{c}{$n_3$} \\ \hline
0.12 &  & 78 & 20 & 2 &  & 234 & 61 & 5 \\
0.25 &  & 56 & 37 & 7 &  & 168 & 113 & 19 \\
0.33 &  & 44 & 44 & 12 &  & 133 & 133 & 34 \\
0.50 &  & 25 & 50 & 25 &  & 75 & 150 & 75 \\ \hline
\end{tabular}
\end{center}
\end{table}

\subsection{Simulation results}
\begin{figure}[p]
\begin{center}
\includegraphics[clip,scale=.25]{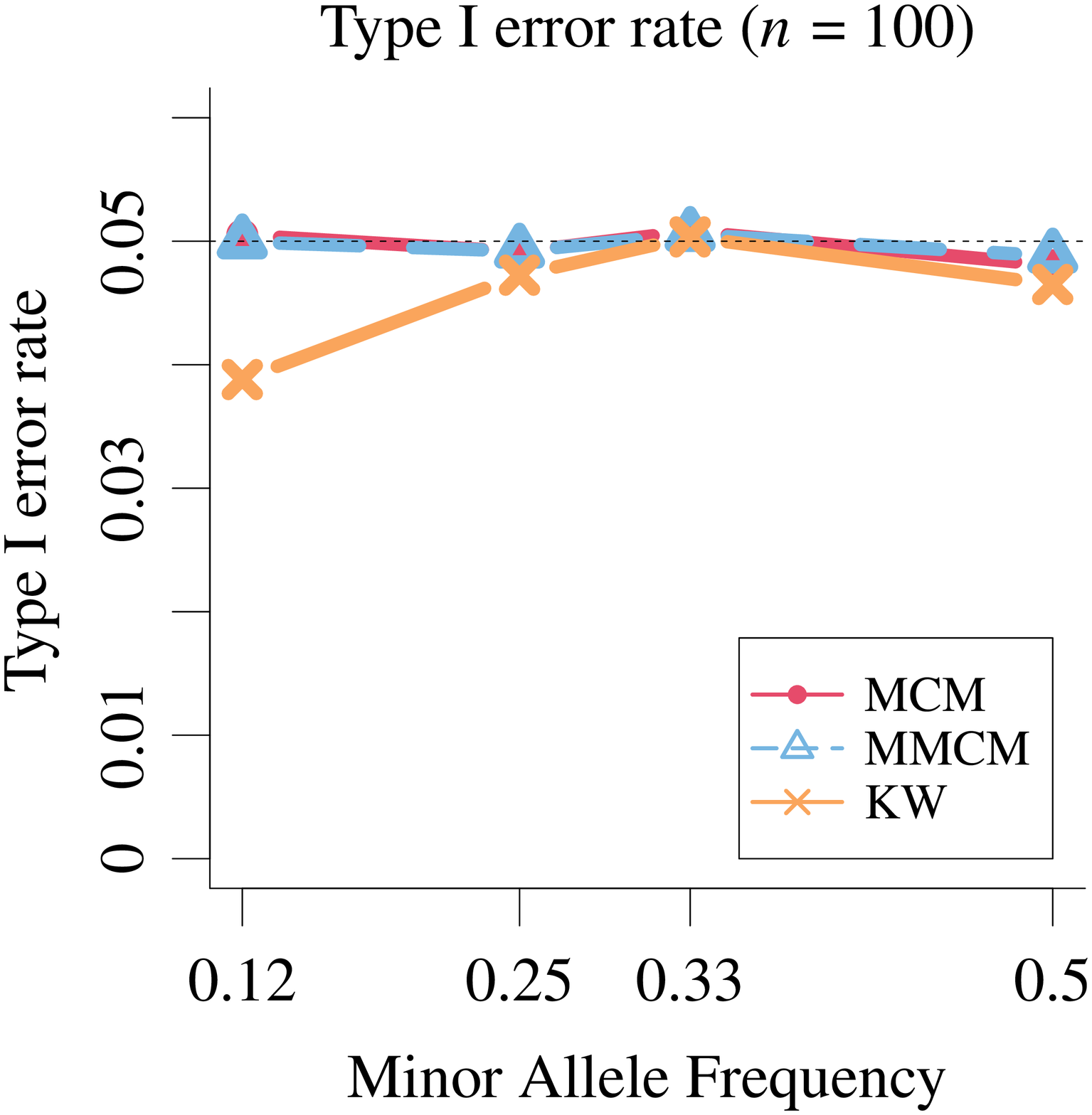}
\includegraphics[clip,scale=.25]{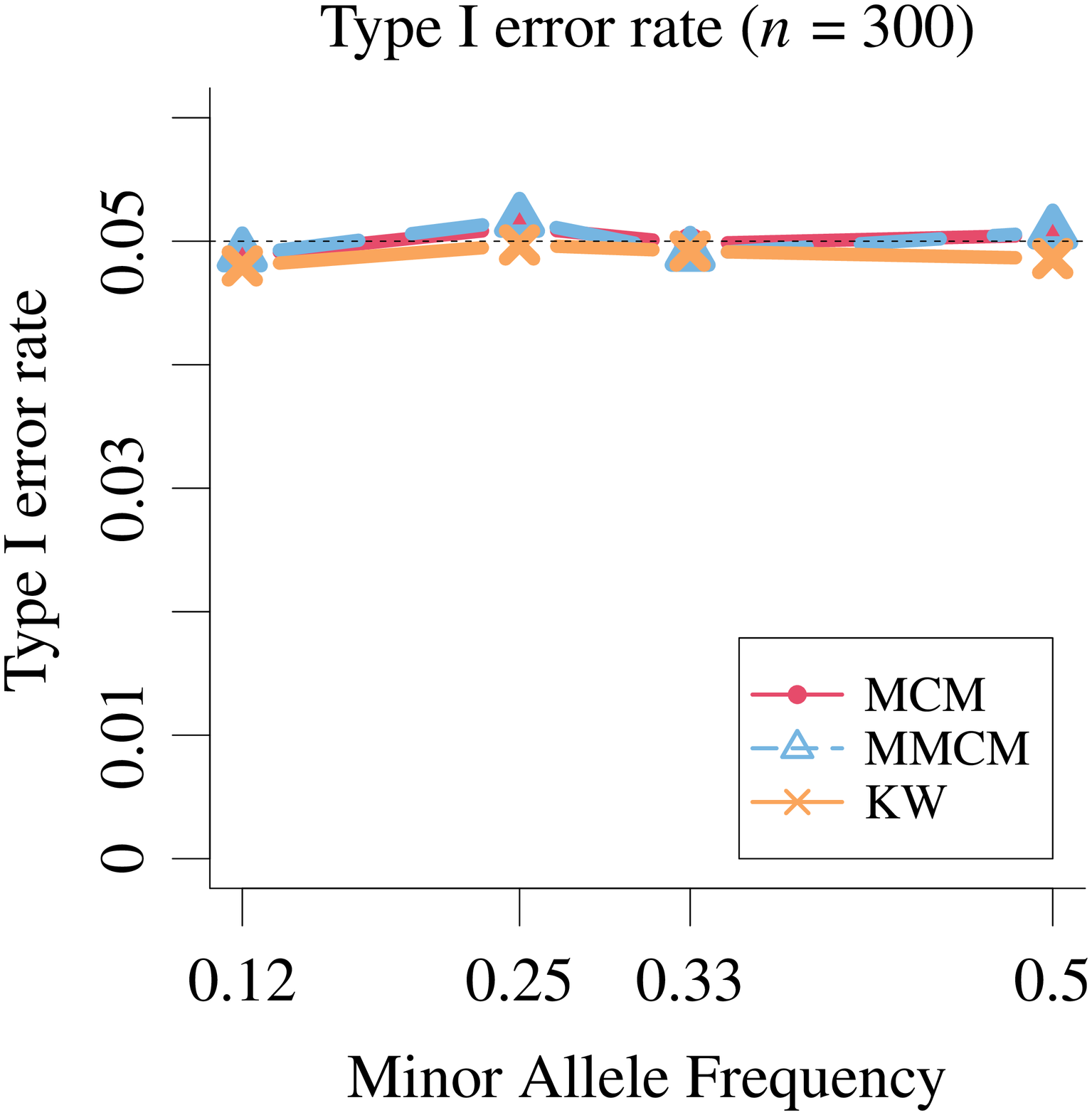}
\caption{
Type I error rates for simulated datasets ($\Delta$ = 0.00).
Abbreviations: MCM, maximum contrast method; MMCM, modified maximum contrast method; KW, Kruskal--Wallis test.
}
\label{fig:alphaerror}
\end{center}
\end{figure}

\begin{figure}[p]
\begin{center}
\includegraphics[clip,scale=.25]{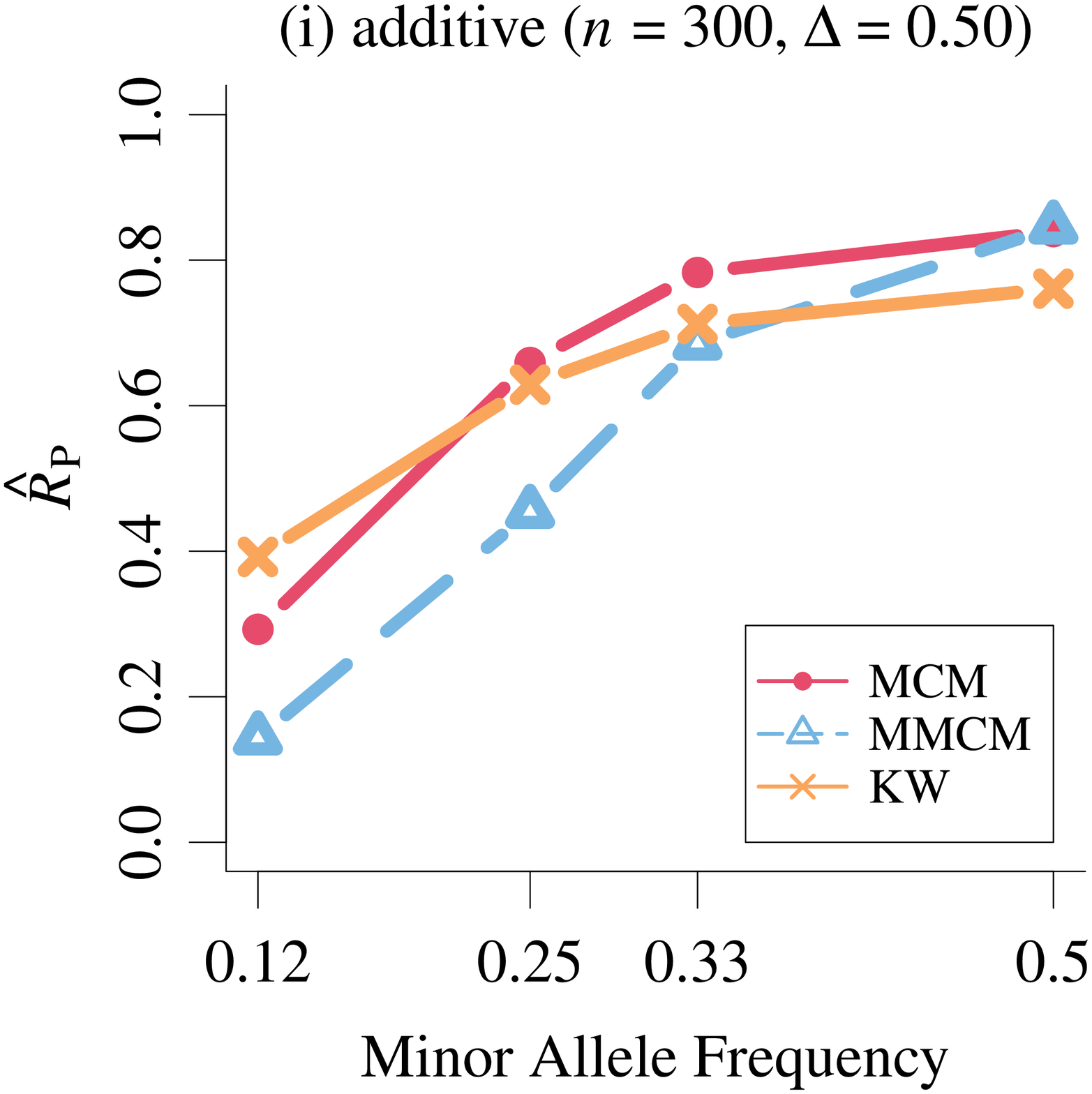}
\includegraphics[clip,scale=.25]{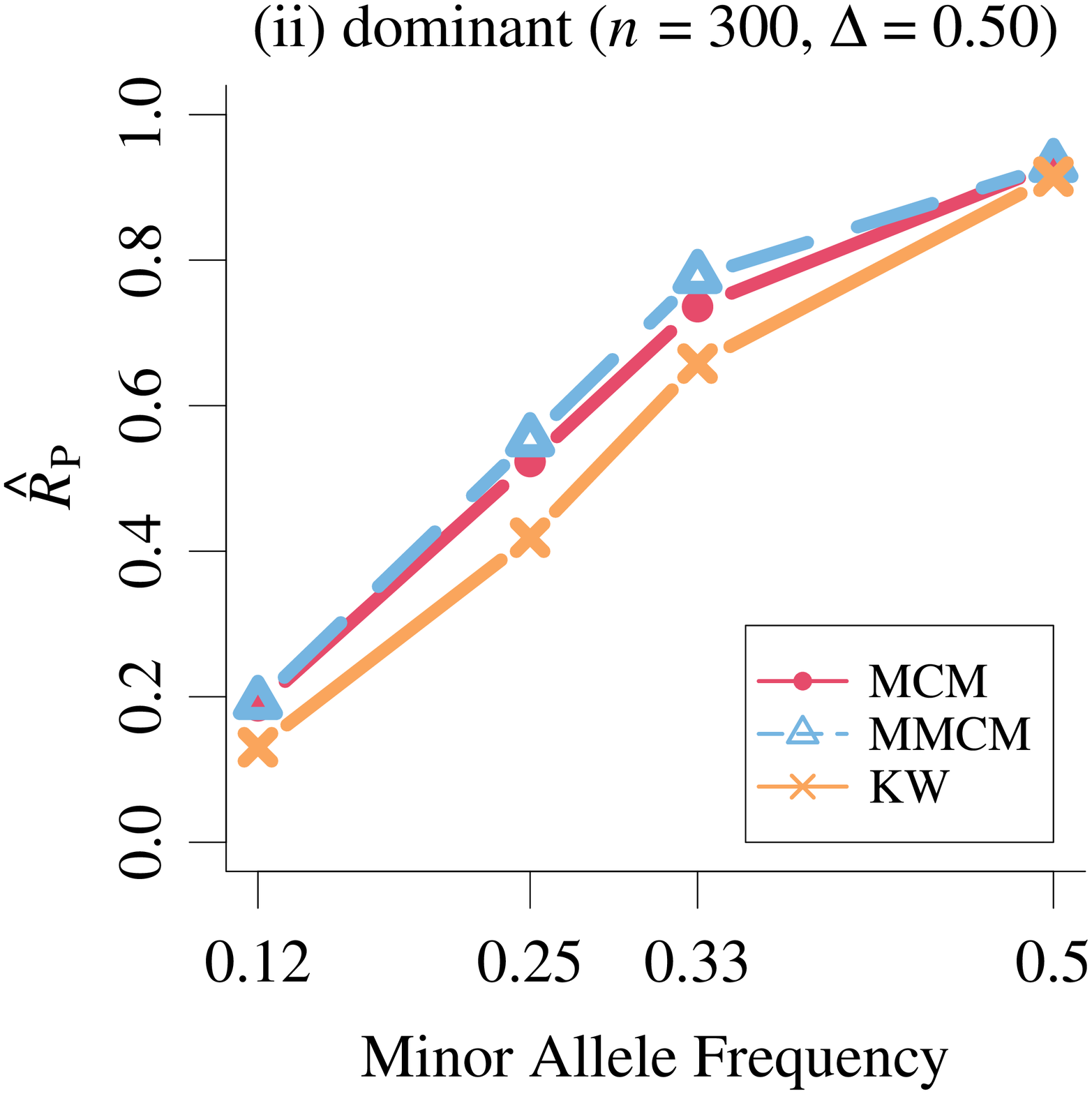}\\ \vspace{.5cm}
\includegraphics[clip,scale=.25]{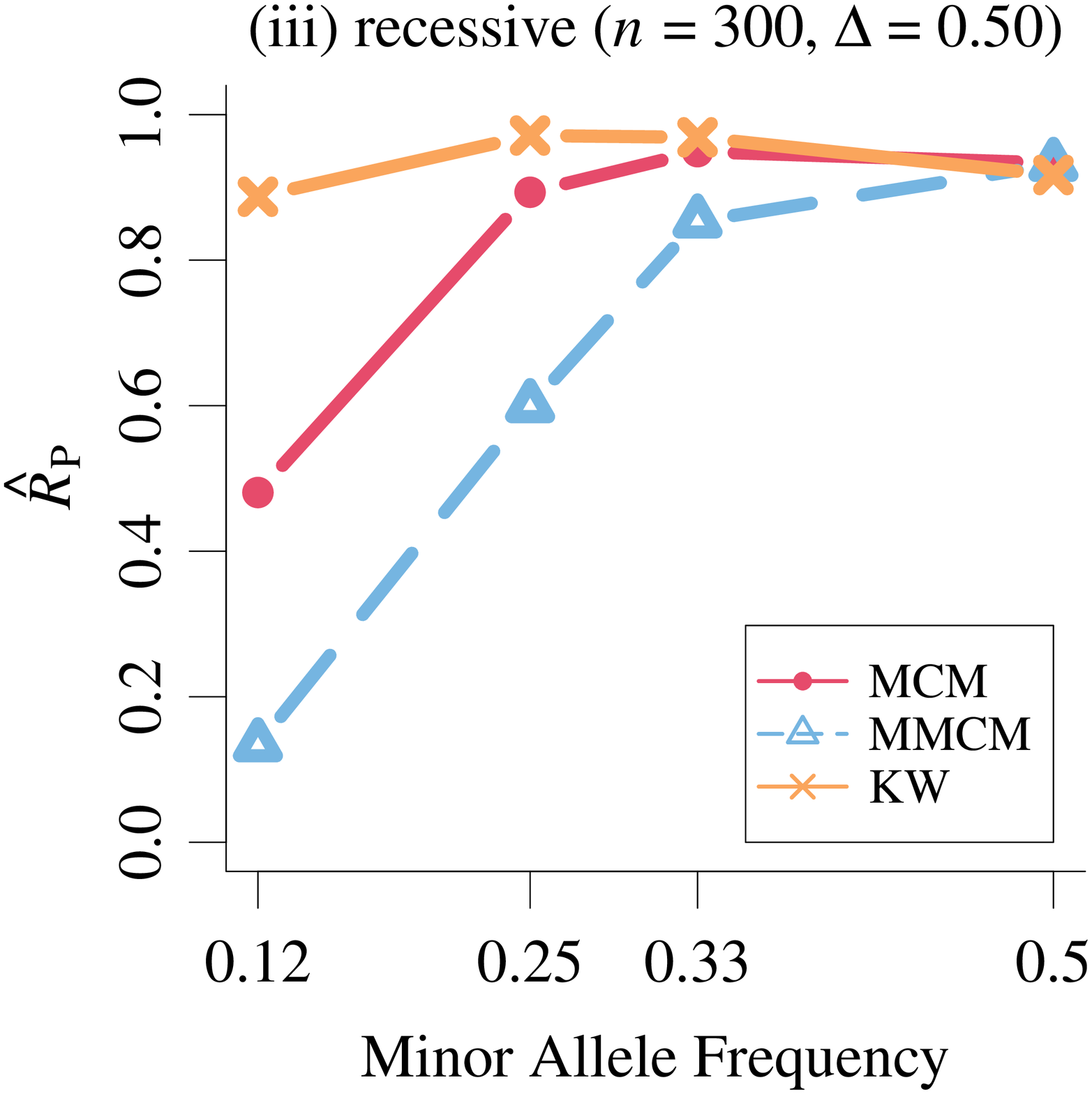}
\caption{
$\hat{R}_{\mathrm{P}}$ (power) for simulated datasets ($n$ = 300, $\Delta$ = 0.50).
Abbreviations: MCM, maximum contrast method; MMCM, modified maximum contrast method; KW, Kruskal--Wallis test.
}
\label{fig:estimateRP}
\end{center}
\end{figure}

\begin{figure}[p]
\begin{center}
\includegraphics[clip,scale=.25]{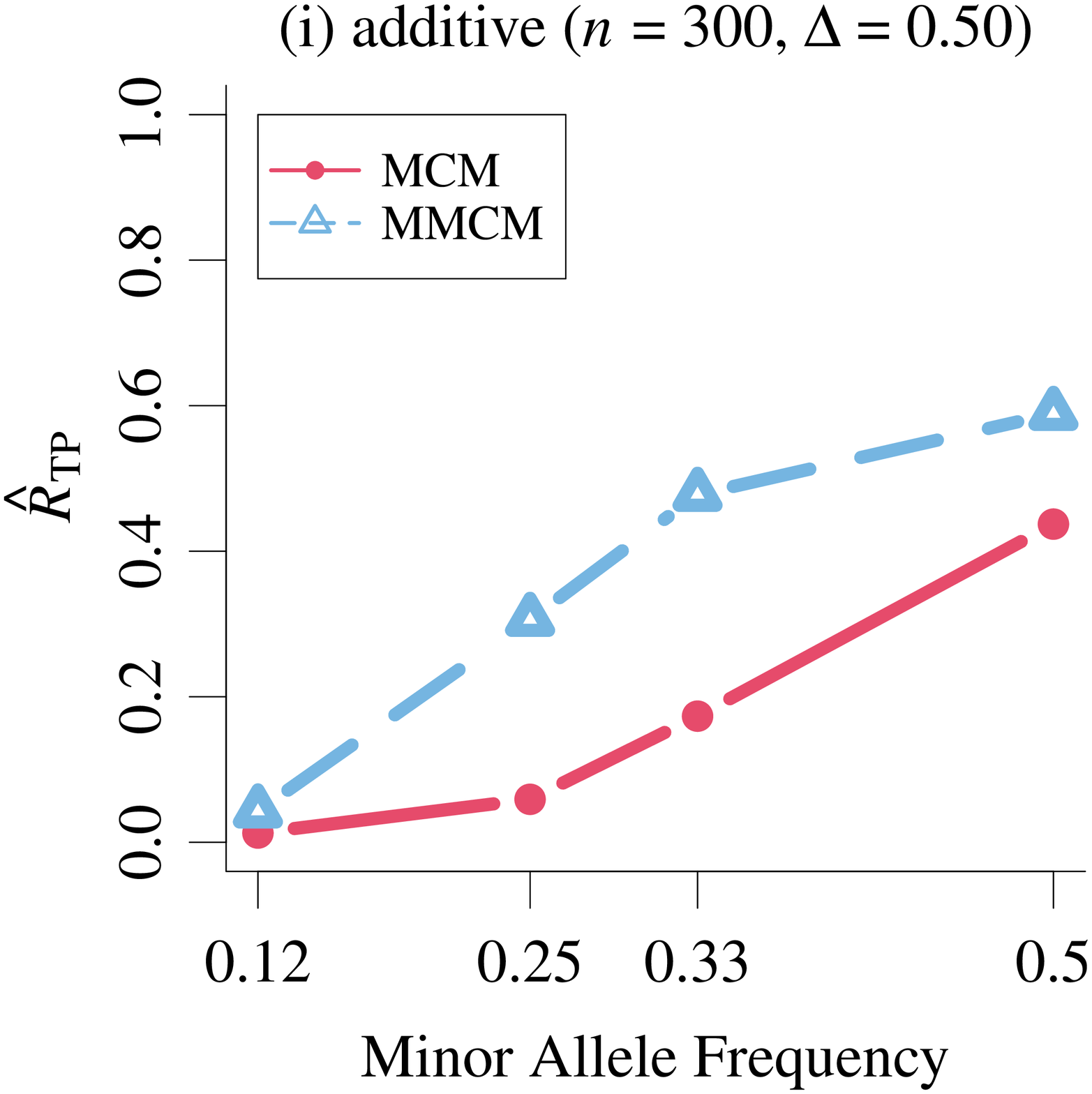}
\includegraphics[clip,scale=.25]{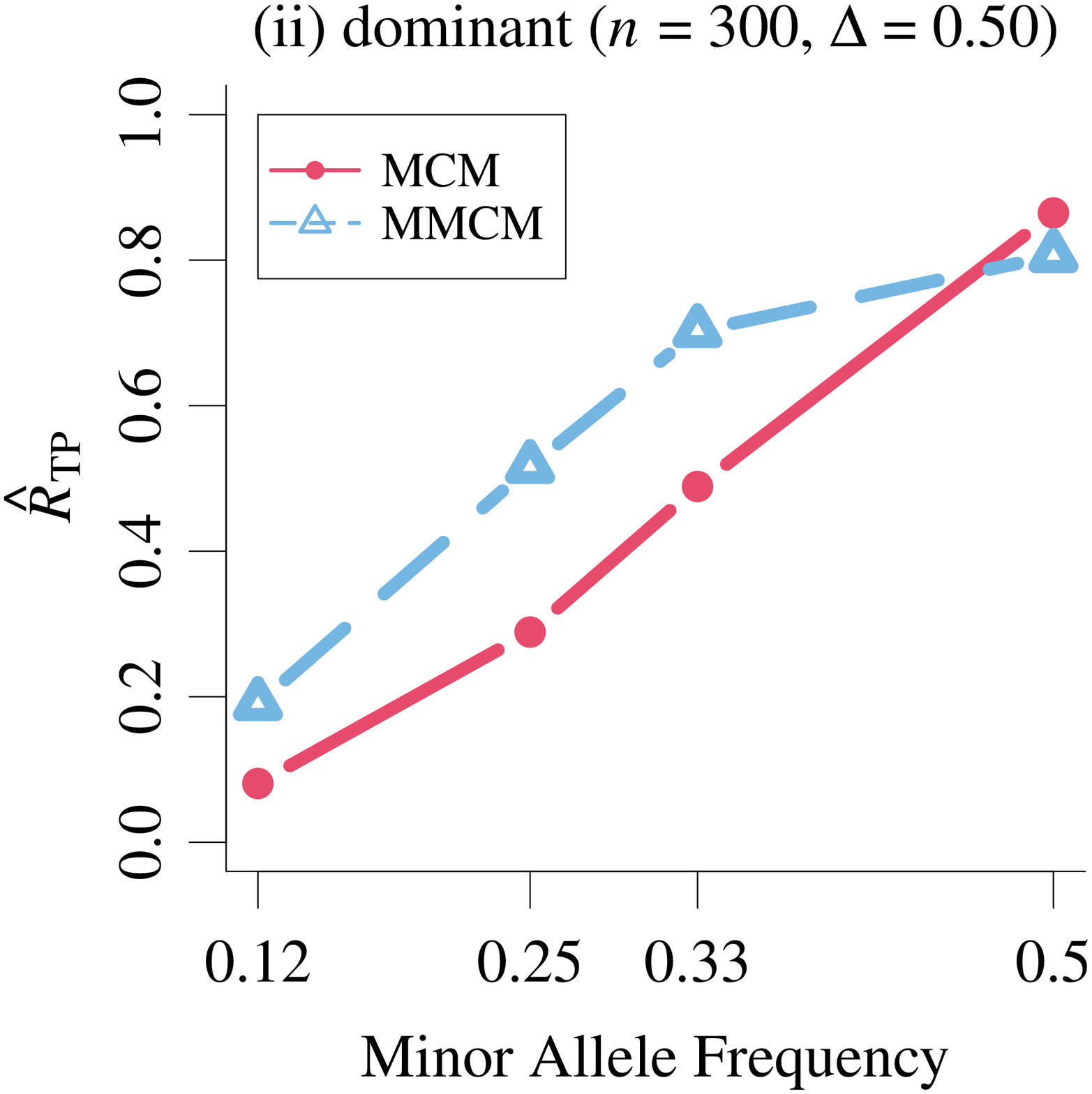}\\ \vspace{.5cm}
\includegraphics[clip,scale=.25]{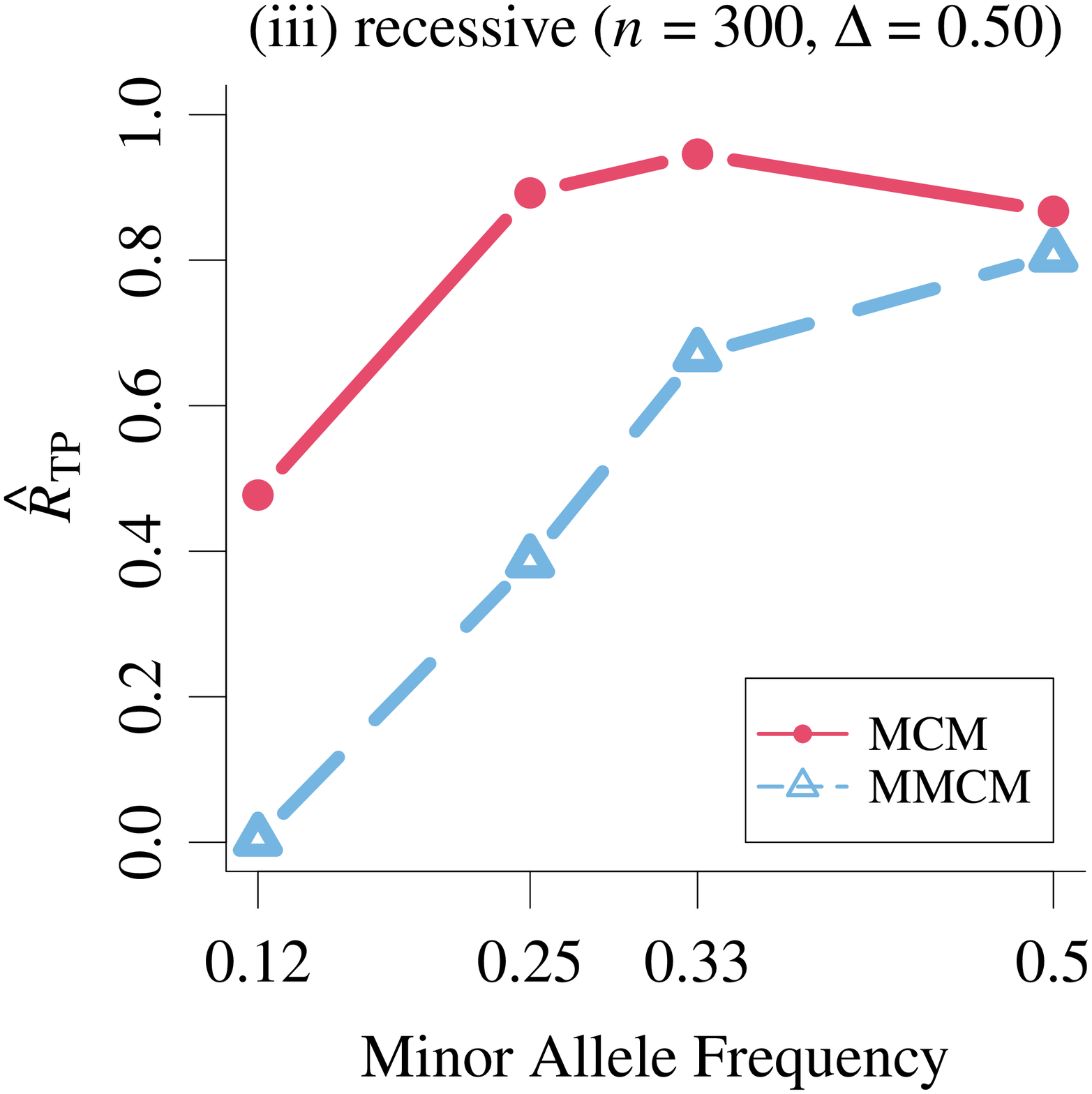}
\caption{
$\hat{R}_{\mathrm{TP}}$ (positive predicted values) for simulated datasets ($n$ = 300, $\Delta$ = 0.50).
Abbreviations: MCM, maximum contrast method; MMCM, modified maximum contrast method; KW, Kruskal--Wallis test.
}
\label{fig:estimateRTP}
\end{center}
\end{figure}

\begin{figure}[p]
\begin{center}
\includegraphics[clip,scale=.25]{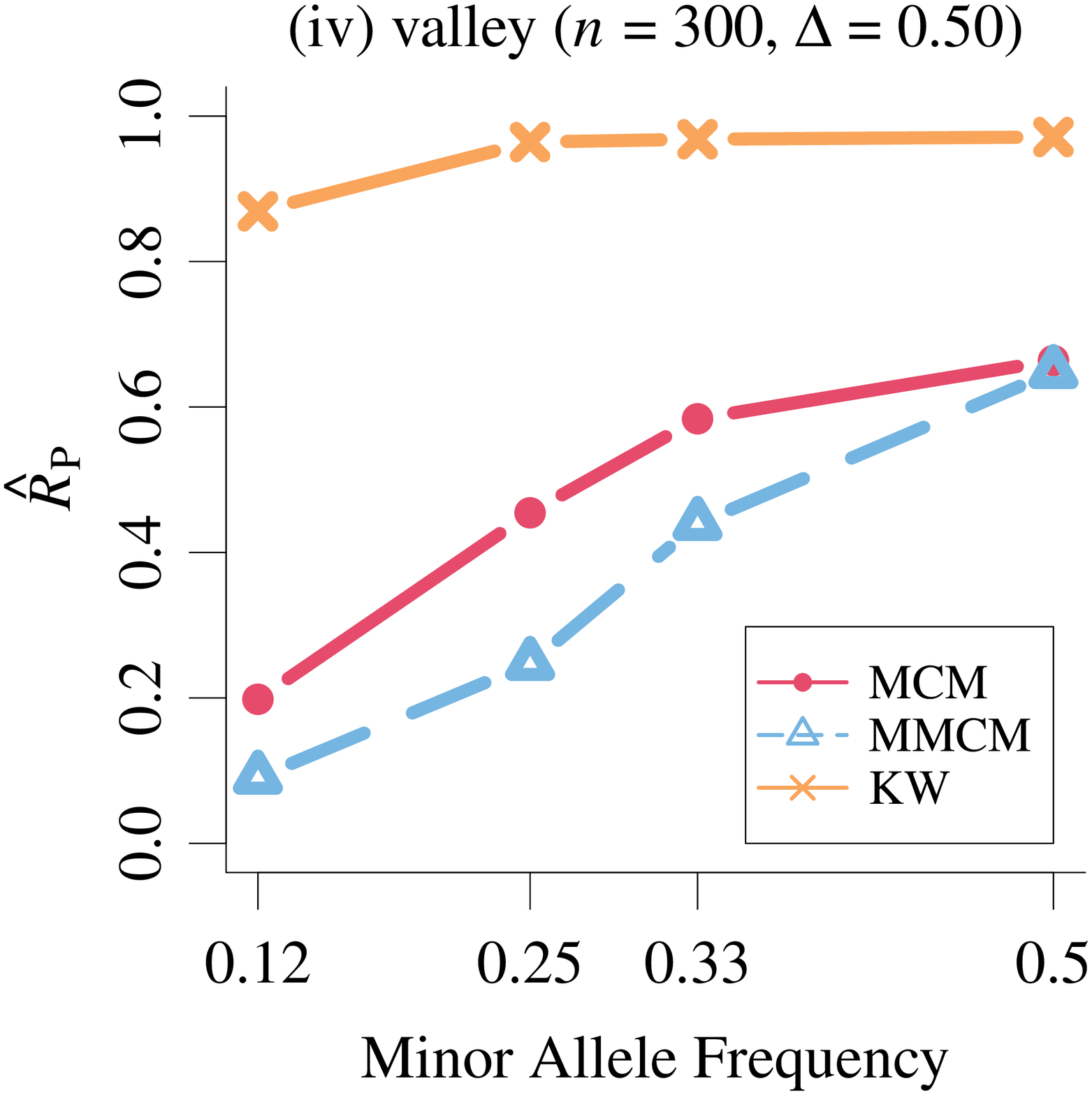}
\caption{
False positive rates for simulated datasets ($n$ = 300, $\Delta$ = 0.50).
Abbreviations: MCM, maximum contrast method; MMCM, modified maximum contrast method.
}
\label{fig:estimateFP}
\end{center}
\end{figure}

The simulation results for each method are shown in Figures \ref{fig:alphaerror}--\ref{fig:estimateFP} for various values of MAF, $\Delta$ = 0.00 or 0.50, and $n$ = 100 or 300.
Further results are given in Supplemental Tables S\ref{tab:AlphaError}--S\ref{tab:FP300}; these results showed the same tendencies discussed below.
The results in Figure \ref{fig:estimateRTP} and Supplemental Tables S\ref{tab:SimResult100} and S\ref{tab:SimResult300} show the positive predictive value ($\hat{R}_{\mathrm{TP}}$) for the detection of the true response patterns.
There is no $\hat{R}_{\mathrm{TP}}$ for the Kruskal--Wallis test because it is an overall test and rejecting the null hypothesis means that there is no difference among genotypes in the population mean of the PK parameters.

The type I error rates (Figure \ref{fig:alphaerror}) were well controlled below the nominal level of 5\% and were below 4\% for the Kruskal--Wallis test at $n$ = 100 and MAF = 0.12.

The $\hat{R}_{\mathrm{P}}$ and $\hat{R}_{\mathrm{TP}}$ increased with increasing $n$ or $\Delta$.
They generally decreased with decreasing MAF.
However, in the results for the maximum contrast method and the Kruskal--Wallis test in the recessive pattern, $\hat{R}_{\mathrm{P}}$ and $\hat{R}_{\mathrm{TP}}$ increased for MAF = 0.5 to 0.33 and decreased for MAF = 0.33 to 0.12 (Figures \ref{fig:estimateRP} and \ref{fig:estimateRTP}, iii), because the balance of the sample size is better at 0.5 than at 0.33 (see Table \ref{tab:MAFandN}).
In contrast, the modified maximum contrast method was robust in unequal-sample-size situations.

For detecting PK-related genes, the $\hat{R}_{\mathrm{P}}$ for the Kruskal--Wallis test was lower than that for the maximum contrast methods, except in the additive pattern (MAF = 0.12) and the recessive pattern (MAF = 0.12, 0.25, and 0.33) (Figure \ref{fig:estimateRP}).
Furthermore, the proportion of false positives was about 0.4--0.6 higher in the Kruskal--Wallis test than in the maximum contrast methods (Figure \ref{fig:estimateFP}) and about 0.01--0.2 lower in the modified maximum contrast method than in the maximum contrast method.
Therefore, the simulation results suggested that the Kruskal--Wallis test detects many SNPs that are not PK-related because this test ignores the order of the response patterns among genotypes.

We evaluated the proportion for detecting the true response pattern in the two maximum contrast methods.
When the MAF was equal to 0.25 or 0.33, in the additive and dominant patterns, the $\hat{R}_{\mathrm{TP}}$ for the modified maximum contrast method was about 0.2--0.3 higher than that for the maximum contrast method (Figure \ref{fig:estimateRTP}, i and ii).
However, in the recessive model, the $\hat{R}_{\mathrm{TP}}$ for the modified maximum contrast method was about 0.5 lower than that for the maximum contrast method (Figure \ref{fig:estimateRTP}, iii).
Therefore, in unequal sample-size situations, the former method was more powerful for detecting the true response pattern in the additive and dominant models, whereas the latter method was more powerful in the recessive model.

\section{Computational speed and accuracy of the modified maximum contrast method}
In this section, we compare the modified maximum contrast methods to assess the computational speed for the same level of accuracy.

\subsection{Simulation conditions}
The simulation conditions were the same as in subsection \ref{sec:simcond}.
\begin{itemize}
\item
Conditions for the pseudo-response values were the same as in subsection \ref{sec:simcond}.
We set five conditions.
\item The total sample size was $n = 300$.
\item The absolute error tolerance of both methods was $10^{-2}$.
\item We evaluated the performance in terms of computational time.
\item Each simulation was repeated 100 times.
\end{itemize}

The methods were implemented in the \texttt{R} language (R-2.10.0) and compiled C and FORTRAN 77 functions.
The modified maximum contrast method was programmed by using the \texttt{R} function \texttt{pmvt()} of Genz and Bretz \cite{GenzBretz1999,GenzBretz2009}.
The simulations were conducted on a personal computer with a 3.0-GHz Intel Core 2 Duo CPU and 3.25 GB of RAM running under 32-bit Windows XP.

\subsection{Simulation results}
The simulation results for each method are given in Table \ref{tab:cmpspeed}.
The permuted modified maximum contrast method required 16.78--298.25 s of computational time.
The computational time for this method was largest for the overall null hypothesis and the valley pattern; these have larger $P$-values than the other cases have.
In contrast, the computational time for the modified maximum contrast method was nearly constant.

\begin{table}[htb]
\begin{center}
\caption{Computational time for each method}
\label{tab:cmpspeed}
\small
\begin{tabular}{lrrrr} \hline
\multicolumn{1}{c}{\multirow{2}{*}{Situation}} & \multicolumn{1}{c}{\multirow{2}{*}{$\Delta$}} & \multicolumn{1}{c}{\multirow{2}{*}{MAF}} & \multicolumn{2}{c}{Sum of computational time (s)} \\ \cline{4-5}
 &  &  & \multicolumn{1}{c}{\phantom{00}pMMCM\phantom{00}} & \multicolumn{1}{c}{MMCM} \\ \hline
overall null hypothesis & 0.00 & 0.33 & 298.25 & 0.92 \\
(i) additive & 0.25 & 0.12 & 94.77 & 0.90 \\
(ii) dominant & 1.00 & 0.50 & 16.78 & 0.92 \\
(iii) recessive & 0.50 & 0.25 & 71.85 & 0.90 \\
(iv) valley & 0.25 & 0.33 & 254.31 & 0.92 \\ \hline
\multicolumn{5}{l}{\scriptsize \shortstack[l]{\\Abbreviations: MAF, minor allele frequency; pMMCM, permuted modified maximum contrast method;\\ MMCM, modified maximum contrast method.}}
\end{tabular}
\end{center}
\end{table}

In genome-wide association studies, 100~000--1~000~000 SNPs are available by using the oligonucleotide SNP array.
Typically, most SNPs have no relation to the PK parameters, and have a large $P$-value; therefore, the simulation results suggested that the modified maximum contrast method is faster than the permuted modified maximum contrast method.

\section{Discussion and recommendations}
In this paper, we proposed the modified maximum contrast method for unequal sample sizes in pharmacogenomic studies.
As this method does not depend on the nuisance parameter, $\sigma^2$, it is not necessary to use approximation.
The use of the randomized quasi-Monte-Carlo method improves the computational speed and accuracy of this method.

Because the modified maximum contrast method is an extension of the permuted modified maximum contrast method, the former method gives similar results for the $P$-value and the best-fit response pattern.
It is however substantially faster and, therefore, a practical choice for unequal sample sizes in large-scale datasets such as those in genome-wide association studies.

The simulation results showed that the modified maximum contrast method is powerful for detecting the true response patterns in the additive and dominant model and has the lowest false-positive rate.
In contrast, the maximum contrast method is powerful for detecting the true response patterns in the recessive model.
The use of a combination of the two methods may be the best approach for screening PK-related genes.

\section{Software}
The modified maximum contrast method is implemented in the \texttt{R} package ``mmcm,'' which is available from CRAN (\url{https://cran.r-project.org/package=mmcm}).
The package also provides the maximum contrast method.

\subsection*{Acknowledgements}
The authors thank the editor and two anonymous reviewers for their constructive comments that helped to greatly improve the presentation of the article.
The authors also thank Prof. Akira Terao for helpful comments.


\clearpage
\appendix

\makeatletter
\def\fnum@table{\tablename\thetable}
\def\fnum@figure{\figurename\thefigure}
\makeatother

\renewcommand\tablename{Supplemental Table S}
\setcounter{table}{0}
\renewcommand\figurename{Supplemental Figure S}
\setcounter{figure}{0}

\section{Numerical examples of the difference between the maximum contrast method and the modified maximum contrast method}
\label{appendix:examplepower}
The following examples illustrate the difference between the maximum contrast method and the modified maximum contrast method.

\begin{figure}[h]
 \begin{minipage}{0.48\hsize}
  \begin{center}
	\includegraphics[width=\textwidth]{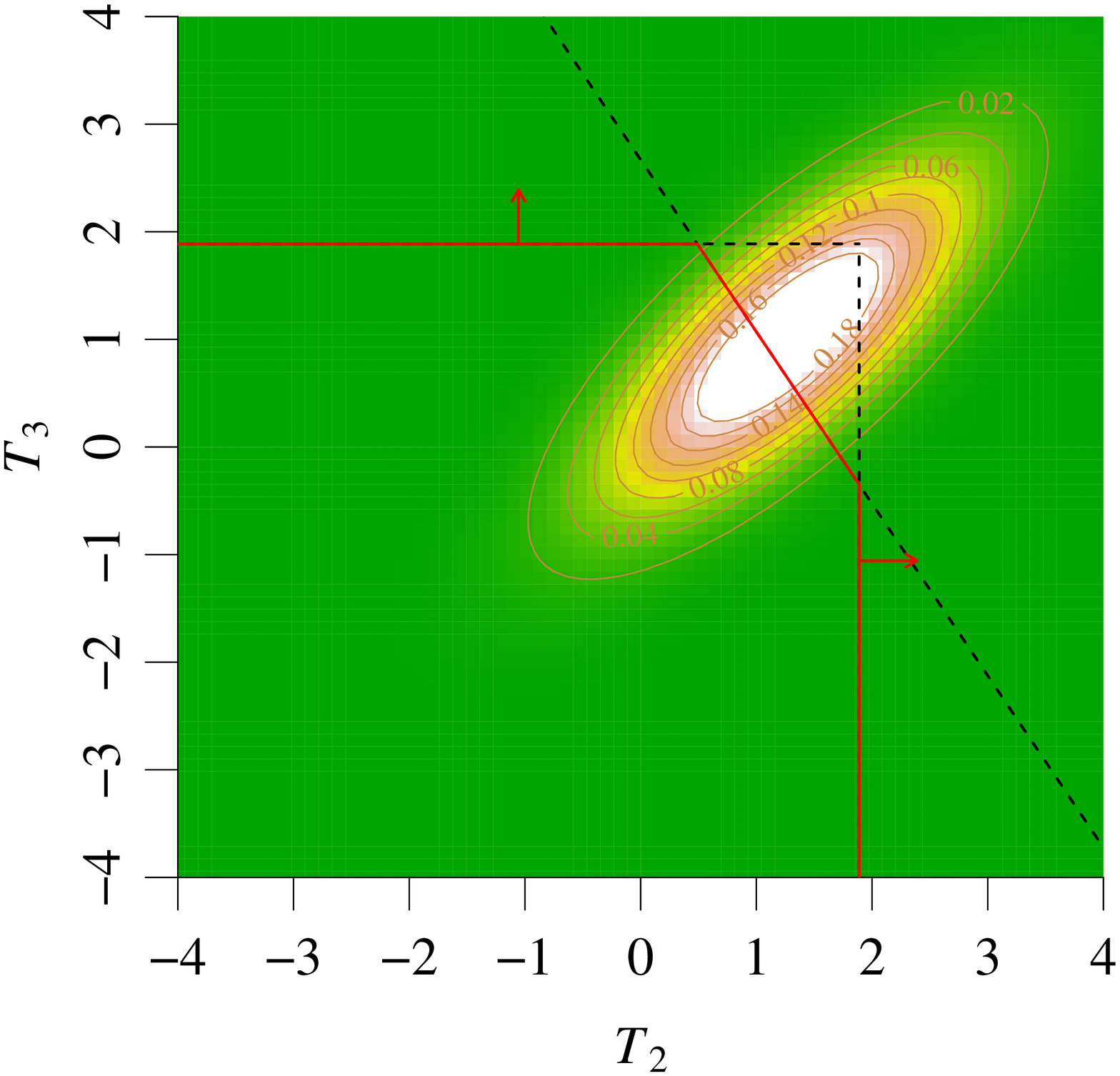}
    \caption{A contour plot and the rejection region of the maximum contrast method;\newline $\beta_T(\g{\mu}=(-1/6, -1/6, 2/6)^{\tp}; \g{C},\g{D})$.}
    \label{fig:one}
  \end{center}
 \end{minipage}
 \hspace{0.02\hsize}
 \begin{minipage}{0.48\hsize}
  \begin{center}
	\includegraphics[width=\textwidth]{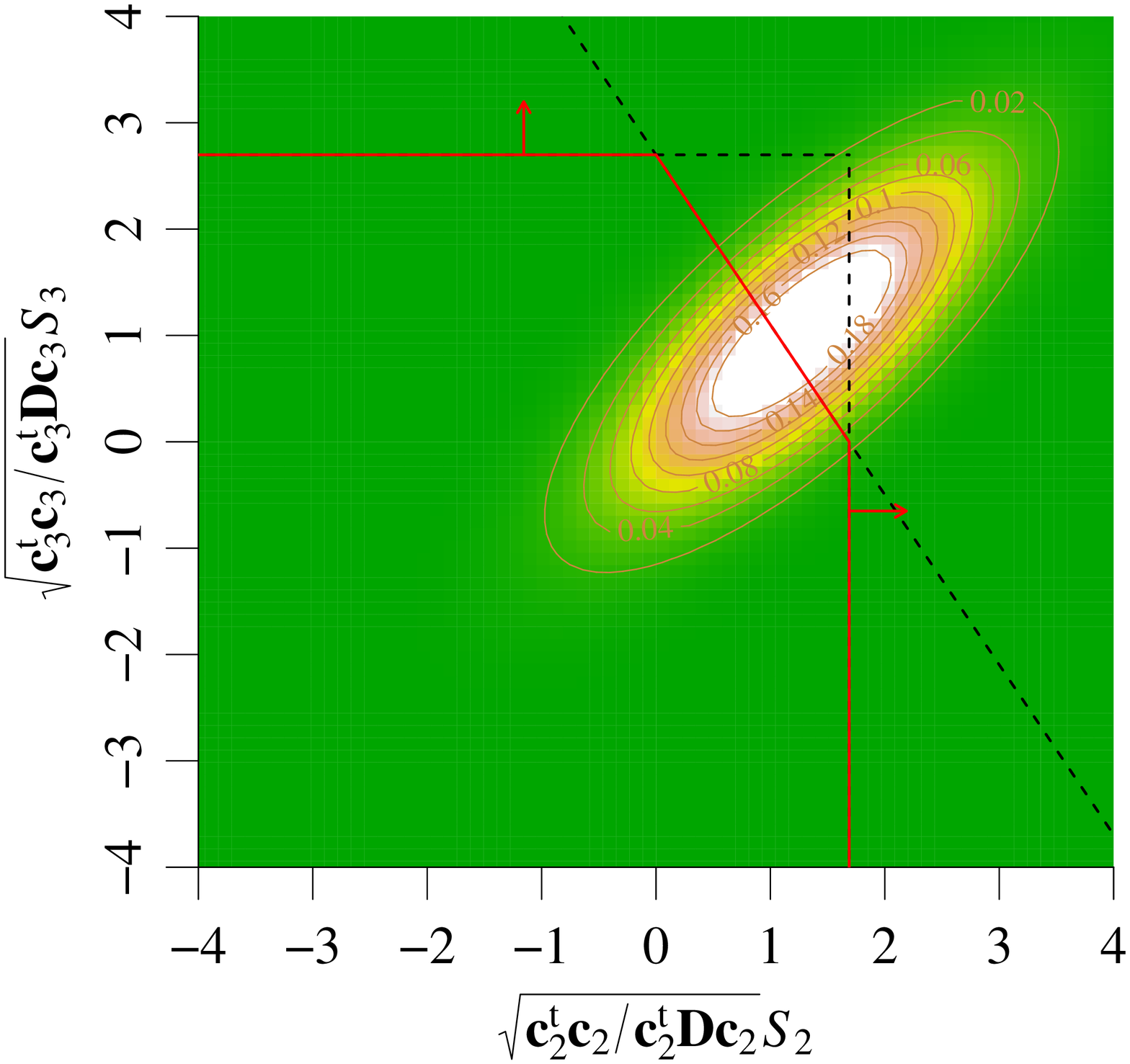}
    \caption{A contour plot and the rejection region of the modified maximum contrast method;\newline $\beta_S(\g{\mu}=(-1/6, -1/6, 2/6)^{\tp}; \g{C},\g{D})$.}
    \label{fig:two}
  \end{center}
 \end{minipage}
\end{figure}

The condition of this numerical example is set to a dominant pattern, taking in account the actual pharmacogenomic studies.
The contrast coefficient matrix is Equation \ref{equ:pgcontrast}, the total sample size is $n=100$, the MAF is 0.25 (($n_1, n_2, n_3$) = (56, 37, 7)), and the significance level is $\alpha=0.05$.
The critical values are
\begin{equation*}
\g{u}_{0.05}=\begin{pmatrix} 1.89 \\ 1.89 \\ 1.89 \end{pmatrix},~~
\g{K}_S^{-1}\g{v}_{0.05}=\begin{pmatrix} 1.91 \\ 1.69 \\ 2.70 \end{pmatrix}.
\end{equation*}
If the true response pattern is a (ii) dominant pattern with $\Delta=0.50$ ($\g{\mu}=(-1/6,\\ -1/6, 2/6)^{\tp}$), then the powers are
\begin{equation}
\begin{split}
\beta_T(\g{\mu}=(-1/6, -1/6, 2/6)^{\tp}; \g{C},\g{D})&=
1 - T_m(-\g{\infty}, \g{u}_{0.05}; \g{\Sigma}_T, \gamma, \g{\lambda}_T)=
0.33,\\
\beta_S(\g{\mu}=(-1/6, -1/6, 2/6)^{\tp}; \g{C},\g{D})&=
1 - T_m(-\g{\infty}, \g{K}_S^{-1}\g{v}_{0.05}; \g{\Sigma}_T, \gamma, \g{\lambda}_T)=
0.35.
\label{equ:powervalue}
\end{split}
\end{equation}
Supplemental Figures S\ref{fig:one} and S\ref{fig:two} show the contour plots and rejection regions of both statistics.
Since the rank of the contrast coefficient matrix $\rank(\g{C})$ equals 2, the statistics $T_1$, $T_2$, and $T_3$ are linearly dependent random variables.
Furthermore, the statistic $T_1$ is expressed by the equation:
\begin{equation*}
T_1=
\frac{\sqrt{\g{c}_2^{\tp}\g{D}\g{c}_2\bigg/\g{c}_2^{\tp}\g{c}_2}}{\sqrt{\g{c}_1^{\tp}\g{D}\g{c}_1\bigg/\g{c}_1^{\tp}\g{c}_1}}T_2+
\frac{\sqrt{\g{c}_3^{\tp}\g{D}\g{c}_3\bigg/\g{c}_3^{\tp}\g{c}_3}}{\sqrt{\g{c}_1^{\tp}\g{D}\g{c}_1\bigg/\g{c}_1^{\tp}\g{c}_1}}T_3.
\end{equation*}

\begin{table}[h]
\begin{center}
\caption{The critical values $\g{K}_S^{-1}\g{v}_{0.05}$ and $\g{u}_{0.05}$ in some situations}
\label{tab:kinvandu}
\begin{tabular}{rcrrrlrrr} \hline
\multicolumn{1}{c}{\multirow{2}{*}{MAF}} & \multirow{2}{*}{Method} & \multicolumn{1}{c}{(i) additive} & \multicolumn{1}{c}{(ii) dominant} & \multicolumn{1}{c}{(iii) recessive} &  & \multicolumn{1}{c}{\multirow{2}{*}{$n_1$}} & \multicolumn{1}{c}{\multirow{2}{*}{$n_2$}} & \multicolumn{1}{c}{\multirow{2}{*}{$n_3$}} \\
 &  & \multicolumn{1}{c}{$k=1$} & \multicolumn{1}{c}{$k=2$} & \multicolumn{1}{c}{$k=3$} &  &  &  &  \\ \cline{1-5} \cline{7-9}
\multirow{2}{*}{0.12} & $\g{u}_{0.05}$ & \cellcolor[gray]{0.9}{1.83} & 1.83 & \cellcolor[gray]{0.9}{1.83} &  & \multirow{2}{*}{78} & \multirow{2}{*}{20} & \multirow{2}{*}{2} \\
 & $\g{K}^{-1}_{S}\g{v}_{0.05}$ & 1.93 & \cellcolor[gray]{0.9}{1.67} & 3.08 &  &  &  &  \\ \cline{1-5} \cline{7-9}
\multirow{2}{*}{0.25} & $\g{u}_{0.05}$ & \cellcolor[gray]{0.9}{1.89} & 1.89 & \cellcolor[gray]{0.9}{1.89} &  & \multirow{2}{*}{56} & \multirow{2}{*}{37} & \multirow{2}{*}{7} \\
 & $\g{K}^{-1}_{S}\g{v}_{0.05}$ & 1.91 & \cellcolor[gray]{0.9}{1.69} & 2.70 &  &  &  &  \\ \cline{1-5} \cline{7-9}
\multirow{2}{*}{0.33} & $\g{u}_{0.05}$ & 1.91 & 1.91 & \cellcolor[gray]{0.9}{1.91} &  & \multirow{2}{*}{44} & \multirow{2}{*}{44} & \multirow{2}{*}{12} \\
 & $\g{K}^{-1}_{S}\g{v}_{0.05}$ & \cellcolor[gray]{0.9}{1.89} & \cellcolor[gray]{0.9}{1.73} & 2.40 &  &  &  &  \\ \cline{1-5} \cline{7-9}
\multirow{2}{*}{0.50} & $\g{u}_{0.05}$ & 1.93 & \cellcolor[gray]{0.9}{1.93} & \cellcolor[gray]{0.9}{1.93} &  & \multirow{2}{*}{25} & \multirow{2}{*}{50} & \multirow{2}{*}{25} \\
 & $\g{K}^{-1}_{S}\g{v}_{0.05}$ & \cellcolor[gray]{0.9}{1.87} & 1.95 & 1.95 &  &  &  &  \\ \hline
\multicolumn{9}{l}{Abbreviation: MAF, minor allele frequency. Shaded region shows the smaller}\\
\multicolumn{9}{l}{critical value from two methods.}\\
\end{tabular}
\end{center}
\end{table}

Thus, it is apparent that the power for the modified maximum contrast method is higher than that for the maximum contrast method, as shown in Supplemental Figures S\ref{fig:one} and S\ref{fig:two}, and in Equation \ref{equ:powervalue}.

Supplemental Table S\ref{tab:kinvandu} shows other examples of the critical values $\g{K}_S^{-1}\g{v}_{0.05}$ and $\g{u}_{0.05}$ with a $\mathrm{MAF} = 0.12, 0.25, 0.33, 0.50$, and $n=100$.
In general, if the element of $\g{K}_S^{-1}\g{v}_{0.05}$ is smaller than the element of $\g{u}_{0.05}$, then the power of the modified maximum contrast method is superior to that of the maximum contrast method in regard to the true response pattern, as shown in Equation \ref{equ:smaxpriority}.
For instance, as shown in Supplemental Table S\ref{tab:kinvandu}, in the case of $\mathrm{MAF} = 0.33$, if the true response pattern is the (i) additive, then the modified maximum contrast method is superior to the maximum contrast method (1.91 vs. 1.89, respectively); if the true response pattern is the (ii) dominant, then the modified maximum contrast method is again superior to the maximum contrast method (1.91 vs. 1.73, respectively); if the true response pattern is the (iii) recessive, then the modified method is instead inferior to the maximum contrast method (1.91 vs. 2.40, respectively).

\begin{figure}[h]
 \begin{minipage}{0.48\hsize}
  \begin{center}
	\includegraphics[width=\textwidth]{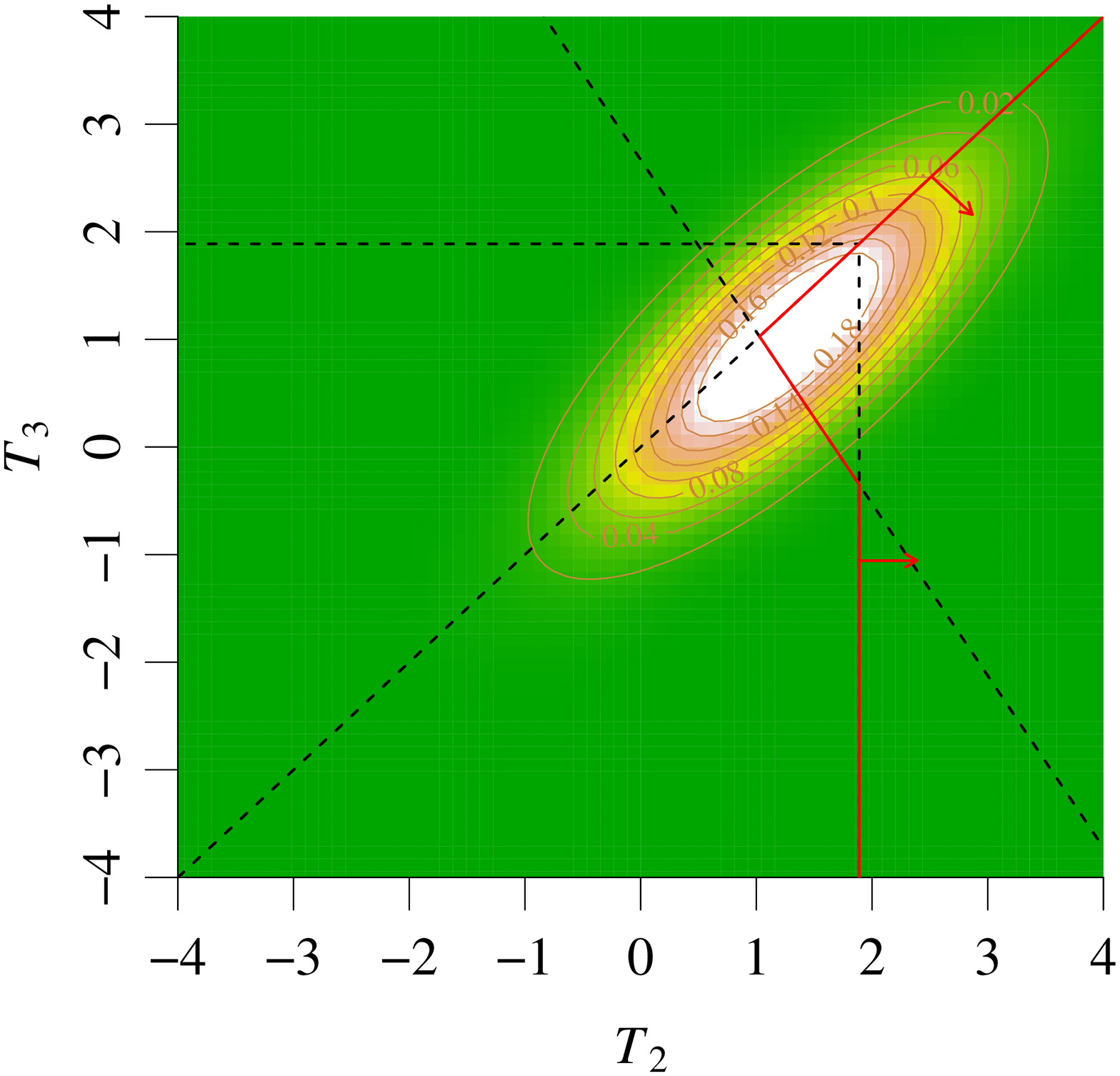}
    \caption{A contour plot and the integrating region of the maximum contrast method;\newline $\Pr(T_{\max} \geq u_{0.05}, T_2 \geq T_1, T_2 \geq T_3 \mid H_1)$}
    \label{fig:onetp}
  \end{center}
 \end{minipage}
 \hspace{0.02\hsize}
 \begin{minipage}{0.48\hsize}
  \begin{center}
	\includegraphics[width=\textwidth]{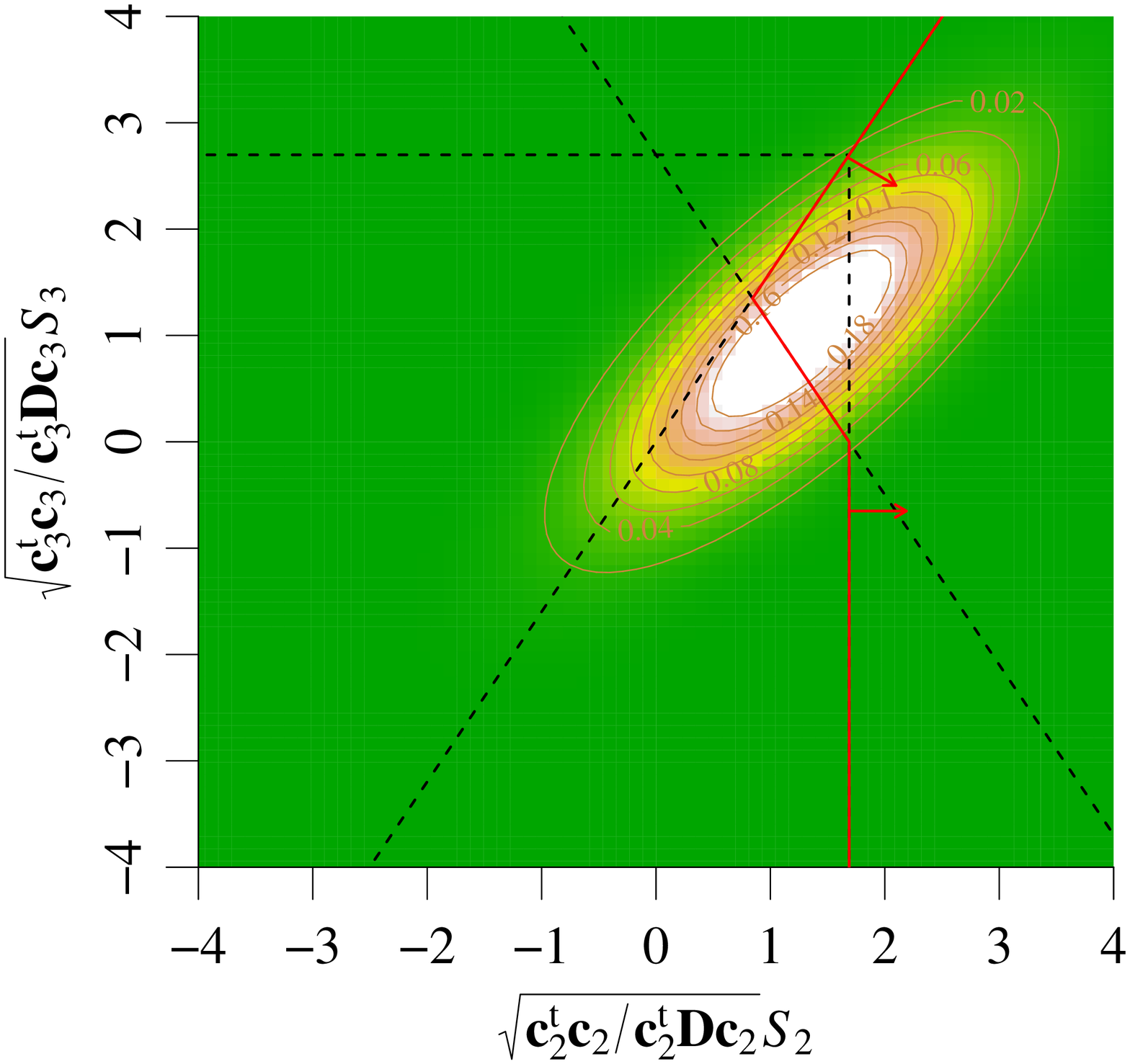}
    \caption{A contour plot and the integrating region of the modi\-fied maximum contrast method;\newline $\Pr(S_{\max} \geq v_{0.05}, S_2 \geq S_1, S_2 \geq S_3 \mid H_1)$}
    \label{fig:twotp}
  \end{center}
 \end{minipage}
\end{figure}

Next, we consider the numerical example of $R_{\mathrm{TP}(T)}$ and $R_{\mathrm{TP}(S)}$ with $n=100$, $\mathrm{MAF}=0.25$, and the (ii) dominant pattern with $\Delta=0.50$ ($\g{\mu}=(-1/6, -1/6, 2/6)^{\tp}$; $k=2$).
From Equation \ref{equ:RTPTS},
\begin{equation}
\begin{split}
R_{\mathrm{TP}(T)}&=
\Pr(T_{\max} \geq u_{0.05}, T_2 \geq T_1, T_2 \geq T_3 \mid H_1)=0.22,\\
R_{\mathrm{TP}(S)}&=
\Pr(S_{\max} \geq v_{0.05}, S_2 \geq S_1, S_2 \geq S_3 \mid H_1)=0.35,
\label{equ:RTPT_RTPS_example}
\end{split}
\end{equation}
using Monte-Carlo integration.
Supplemental Figures S\ref{fig:onetp} and S\ref{fig:twotp} show contour plots and integrating regions of both statistics.

\begin{table}[h]
\begin{center}
\small
\caption{The relationship between true response patterns and non-central parameter vectors}
\label{tab:vecandparm}
\begin{tabular}{clrrr} \hline
\multirow{2}{*}{MAF} & \multirow{2}{*}{True response pattern} & \multicolumn{3}{c}{non-central parameter vector ($\g{\lambda}_T$)} \\ \cline{3-5}
 &  & \multicolumn{1}{c}{(i) additive} & \multicolumn{1}{c}{(ii) dominant} & \multicolumn{1}{c}{(iii) recessive} \\ \hline
\multirow{3}{*}{0.12} & (i) additive & 0.70 & 0.52 & \cellcolor[gray]{0.9}{0.97} \\
 & (ii) dominant & \cellcolor[gray]{0.9}{0.70} & \cellcolor[gray]{0.9}{0.70} & 0.64 \\
 & (iii) recessive & 0.70 & 0.35 & \cellcolor[gray]{0.9}{1.29} \\ \hline
\multirow{3}{*}{0.25} & (i) additive & 1.25 & 0.96 & \cellcolor[gray]{0.9}{1.53} \\
 & (ii) dominant & 1.25 & \cellcolor[gray]{0.9}{1.27} & 1.02 \\
 & (iii) recessive & 1.25 & 0.64 & \cellcolor[gray]{0.9}{2.04} \\ \hline
\multirow{3}{*}{0.33} & (i) additive & 1.54 & 1.22 & \cellcolor[gray]{0.9}{1.69} \\
 & (ii) dominant & 1.54 & \cellcolor[gray]{0.9}{1.62} & 1.13 \\
 & (iii) recessive & 1.54 & 0.81 & \cellcolor[gray]{0.9}{2.25} \\ \hline
\multirow{3}{*}{0.50} & (i) additive & \cellcolor[gray]{0.9}{1.77} & 1.60 & 1.60 \\
 & (ii) dominant & 1.77 & \cellcolor[gray]{0.9}{2.13} & 1.07 \\
 & (iii) recessive & 1.77 & 1.07 & \cellcolor[gray]{0.9}{2.13} \\ \hline
\multicolumn{5}{l}{\shortstack[l]{\\Shaded region shows the highest absolute value of the element of the\\ non-central parameter vector.}}
\end{tabular}
\end{center}
\end{table}

Thus, $R_{\mathrm{TP}(S)}$ is higher than $R_{\mathrm{TP}(T)}$, as shown in Supplemental Figures S\ref{fig:onetp} and S\ref{fig:twotp}, and in Equation \ref{equ:RTPT_RTPS_example}.

Supplemental Table S\ref{tab:vecandparm} shows the relationship between the true response patterns and the non-central parameter vectors.
We considered true response patterns that were (i) additive, (ii) dominant pattern, and (iii) recessive with $\Delta=0.50$ and a $\mathrm{MAF} = 0.12, 0.25, 0.33$, and 0.50.

The element of the non-contral parameter vector that corresponds to the true response pattern must be the highest value in order to achieve a high $R_{\mathrm{TP}(T)}$; however, as shown in Supplemental Table S\ref{tab:vecandparm}, this requirement is not satisfied when the MAF equals 0.12, 0.25, or 0.33. For example, in the case of $\mathrm{MAF} = 0.33$, if the true response pattern is (i) additive, then the element of the (iii) recessive pattern is the highest value (0.97).

On the other hand, the modified maximum contrast method gives priority to $S_1$ ((i) additive) or $S_2$ ((ii) dominant) when the MAF is equal to 0.12, 0.25, or 0.33, as shown in Supplemental Table S\ref{tab:kinvandu}.
In other words, this method adjusts the lower value of the non-contral parameter shown in Supplemental Table S\ref{tab:vecandparm} ((i) additive and (ii) dominant).
Therefore, $R_{\mathrm{TP}(S)}$ is expected to be higher than $R_{\mathrm{TP}(T)}$ with the (i) additive and (ii) dominant pattern in the conditions that are similar to those of actual pharmacogenomic studies.

\section{The simultaneous distribution of the proposed statistic}
\label{appendix:derive}
Now, we derive the simultaneous distribution of $\g{S} = (S_1,\, S_2,\, \ldots,\, S_k,\, \ldots, S_m)^{\tp}$ and show that $\g{S}$ does not depend on $\sigma^2$.
As $Z_k^{\prime} \sim N(\g{c}_k^{\tp}\g{\mu}\Big/\sqrt{\sigma^2\g{c}_k^{\tp}\g{c}_k},\,\g{c}_k^{\tp}\g{D}\g{c}_k\Big/\g{c}_k^{\tp}\g{c}_k)$, $\g{Z}^{\prime}=(Z_1^{\prime},\, Z_2^{\prime},\, \ldots,\, Z_k^{\prime},\, \ldots,\, Z_m^{\prime})^{\tp}$ follows $m$-variate normal distribution $N_m(\g{\lambda}_S, \g{\Sigma}_S)$, where $\g{\lambda}_S=\left\{\g{c}_k^{\tp}\g{\mu}\Big/\sqrt{\sigma^2\g{c}_k^{\tp}\g{c}_k}\right\}_{1\leq k\leq m}$ is population mean vector and either singular or non-singular covariance matrix is
\begin{equation*}
\g{\Sigma}_S=
\left\{
\frac{\g{c}_k^{\tp}\g{D}\g{c}_l}{\sqrt{\g{c}_k^{\tp}\g{c}_k}\sqrt{\g{c}_k^{\tp}\g{c}_l}}
\right\}_{1\leq k,l \leq, m}
~.
\end{equation*}
Statistic $\g{S}$ can be written as $\g{S}=\g{Z}^{\prime}\big/\sqrt{W/\gamma}$ according to Equation \ref{equ:smax}, where $W=\gamma V\big/\sigma^2$ is independently $\g{Z}^{\prime}$ and follows a chi-square distribution with $\gamma$ degrees of freedom.
Thus, the joint distribution of $(\g{Z}^{\prime}, W)$ is
\begin{equation}
\begin{split}
f(\g{Z}^{\prime}&=\g{z}^{\prime}, W=w \mid \g{\Sigma}_S, \gamma, \g{\lambda}_S)\\&=
\frac{|\g{\Sigma}_S|^{-\frac{1}{2}}}{2^{\frac{m+\gamma}{2}}\pi^{\frac{m}{2}}\Gamma\left(\frac{\gamma}{2}\right)}\exp\left[
-\frac{1}{2}\left\{
\left(\g{z}^{\prime}-\g{\lambda}_S \right)^{\mathrm t}
\g{\Sigma}_S^{-1}
\left(\g{z}^{\prime}-\g{\lambda}_S \right)
+w
\right\}\right]
w^{\frac{\gamma}{2}-1}~.
\end{split}
\label{equ:JointZW}
\end{equation}
Consider the transformation $\g{S}=\g{Z}^{\prime}\big/\sqrt{W/\gamma}, W=W$ and the inverse transformation $\g{Z}^{\prime}=\g{S}\sqrt{W/\gamma}, W=W$.
The Jacobian of the transformation is
\begin{equation*}
|J|=
\left|
\begin{array}{ccccc}
\sqrt{w/\gamma} & 0 & \cdots & 0 & * \\
0 & \sqrt{w/\gamma} & \cdots & 0 & * \\
\vdots & \vdots & \ddots & 0 & * \\
0 & 0 & \cdots & \sqrt{w/\gamma} & * \\
0 & 0 & \cdots & 0 & 1
\end{array}
\right|=
(w/\gamma)^{\frac{m}{2}}~.
\end{equation*}
Note that $*$ is equal to 0 in the calculation.
Thus, the joint distribution of $(\g{S}, W)$ is
\begin{equation}
f(\g{S}=\g{s}, W=w \mid \g{\Sigma}_S, \gamma, \g{\lambda}_S) = f(\g{Z}^{\prime}=\g{s}\sqrt{w/\gamma}, W=w \mid \g{\Sigma}_S, \gamma, \g{\lambda}_S) \times ||J||.
\label{equ:distSW}
\end{equation}
Thus, the distribution of $\g{S}$ is obtained from Equations \ref{equ:JointZW} and \ref{equ:distSW} by integrating out $w$,
\begin{equation}
\begin{split}
&f(\g{S}=\g{s} \mid \g{\Sigma}_S, \gamma, \g{\lambda}_S)=
\int_{0}^{+\infty}f(\g{S}=\g{s}, W=w \mid \g{\Sigma}_S, \gamma, \g{\lambda}_S) \idif w\\&=
\frac{\left(\frac{1}{2}\right)^{\frac{m+\gamma}{2}}|\g{\Sigma}_S|^{-\frac{1}{2}}}{(\pi\gamma)^{\frac{m}{2}}\Gamma\left(\frac{\gamma}{2}\right)}\times\\&
\int_0^{+\infty}
\exp\left[
-\frac{1}{2}\left\{
\left(\g{s}\sqrt{w/\gamma}-\g{\lambda}_S \right)^{\mathrm t}
\g{\Sigma}_S^{-1}
\left(\g{s}\sqrt{w/\gamma}-\g{\lambda}_S \right)
+w
\right\}\right]
w^{\frac{m+\gamma}{2}-1}
\idif w~.
\end{split}
\label{equ:densityS}
\end{equation}
Finally, under the overall null hypothesis, $\g{c}_k^{\tp}\g{\mu}=0$ holds, then $\g{\lambda}_S=\g{0}$ holds.
Thus, Equation \ref{equ:densityS} becomes
\begin{equation*}
\begin{split}
f(\g{S}&=\g{s}\mid \g{\Sigma}_S, \gamma)
\\&=
\frac{\left(\frac{1}{2}\right)^{\frac{m+\gamma}{2}}|\g{\Sigma}_S|^{-\frac{1}{2}}}{(\pi\gamma)^{\frac{m}{2}}\Gamma\left(\frac{\gamma}{2}\right)}
\int_0^{+\infty}
\exp\left[
-\frac{1}{2}\left\{
\frac{\g{s}^{\mathrm t}\g{\Sigma}_S^{-1}\g{s}}{\gamma}+1
\right\}w\right]
w^{\frac{m+\gamma}{2}-1}
\idif w
\\&=
\frac{\left(\frac{1}{2}\right)^{\frac{m+\gamma}{2}}|\g{\Sigma}_S|^{-\frac{1}{2}}}{(\pi\gamma)^{\frac{m}{2}}\Gamma\left(\frac{\gamma}{2}\right)}
\left\{
\frac{1}{2}\left(
\frac{\g{s}^{\mathrm t}\g{\Sigma}_S^{-1}\g{s}}{\gamma}+1
\right)\right\}^{-\frac{m+\gamma}{2}}\Gamma\left(\frac{m+\gamma}{2}\right)
\\&=
\frac{|\g{\Sigma}_S|^{-\frac{1}{2}}\Gamma\left(\frac{m+\gamma}{2}\right)}{(\pi\gamma)^{\frac{m}{2}}\Gamma\left(\frac{\gamma}{2}\right)}
\left(
\frac{\g{s}^{\mathrm t}\g{\Sigma}_S^{-1}\g{s}}{\gamma}+1
\right)^{-\frac{m+\gamma}{2}}
\end{split}~.
\end{equation*}
Therefore, $\g{S}$ follows $m$-variate $t$-distribution $t_m(\g{\Sigma}_S, \gamma)$ under the overall null hypothesis.

\section{Supplemental tables}

\begin{table}[hb]
\begin{center}
\caption{Type I error rates}
\label{tab:AlphaError}
\small
\begin{tabular}{ccrr} \hline
MAF & Method & \multicolumn{1}{c}{$n = 100$} & \multicolumn{1}{c}{$n = 300$} \\ \hline
\multirow{3}{*}{0.12} & MCM & 0.050 & 0.049 \\
 & MMCM & 0.051 & 0.049 \\
 & KW & 0.039 & 0.048 \\ \hline
\multirow{3}{*}{0.25} & MCM & 0.049 & 0.052 \\
 & MMCM & 0.049 & 0.051 \\
 & KW & 0.047 & 0.050 \\ \hline
\multirow{3}{*}{0.33} & MCM & 0.051 & 0.049 \\
 & MMCM & 0.051 & 0.050 \\
 & KW & 0.050 & 0.049 \\ \hline
\multirow{3}{*}{0.50} & MCM & 0.049 & 0.051 \\
 & MMCM & 0.048 & 0.051 \\
 & KW & 0.047 & 0.049 \\ \hline
\multicolumn{4}{l}{\scriptsize \shortstack[l]{\\Abbreviations: MAF, minor allele frequency; MCM,\\ maximum contrast method; MMCM, modified maxi- \\ mum contrast method; KW, Kruskal--Wallis test.}}
\end{tabular}
\end{center}
\end{table} %

\begin{landscape}

\begin{table}[p]
\begin{center}
\vspace{-10pt}
\caption{\small The $\hat{R}_{\mathrm{P}}$ (power) and $\hat{R}_{\mathrm{TP}}$ (positive predicted value) for various response patterns $(n = 100)$}
\label{tab:SimResult100}
\scriptsize
\begin{tabular}{cccrrrrrrrrrrrr} \hline
 &  &  & \multicolumn{4}{c}{$\Delta$ = 0.25} & \multicolumn{4}{c}{$\Delta$ = 0.50} & \multicolumn{4}{c}{$\Delta$ = 1.00} \\ \hhline{---------------}
MAF & True situation & Method & \multicolumn{1}{c}{(i)} & \multicolumn{1}{c}{(ii)} & \multicolumn{1}{c}{(iii)} & \multicolumn{1}{c}{$\hat{R}_{\mathrm{P}}$} & \multicolumn{1}{c}{(i)} & \multicolumn{1}{c}{(ii)} & \multicolumn{1}{c}{(iii)} & \multicolumn{1}{c}{$\hat{R}_{\mathrm{P}}$} & \multicolumn{1}{c}{(i)} & \multicolumn{1}{c}{(ii)} & \multicolumn{1}{c}{(iii)} & \multicolumn{1}{c}{$\hat{R}_{\mathrm{P}}$} \\ \hhline{---------------}
\multirow{9}{*}{0.12} & \multirow{3}{*}{\shortstack{(i) additive\\ \includegraphics[width=.5cm]{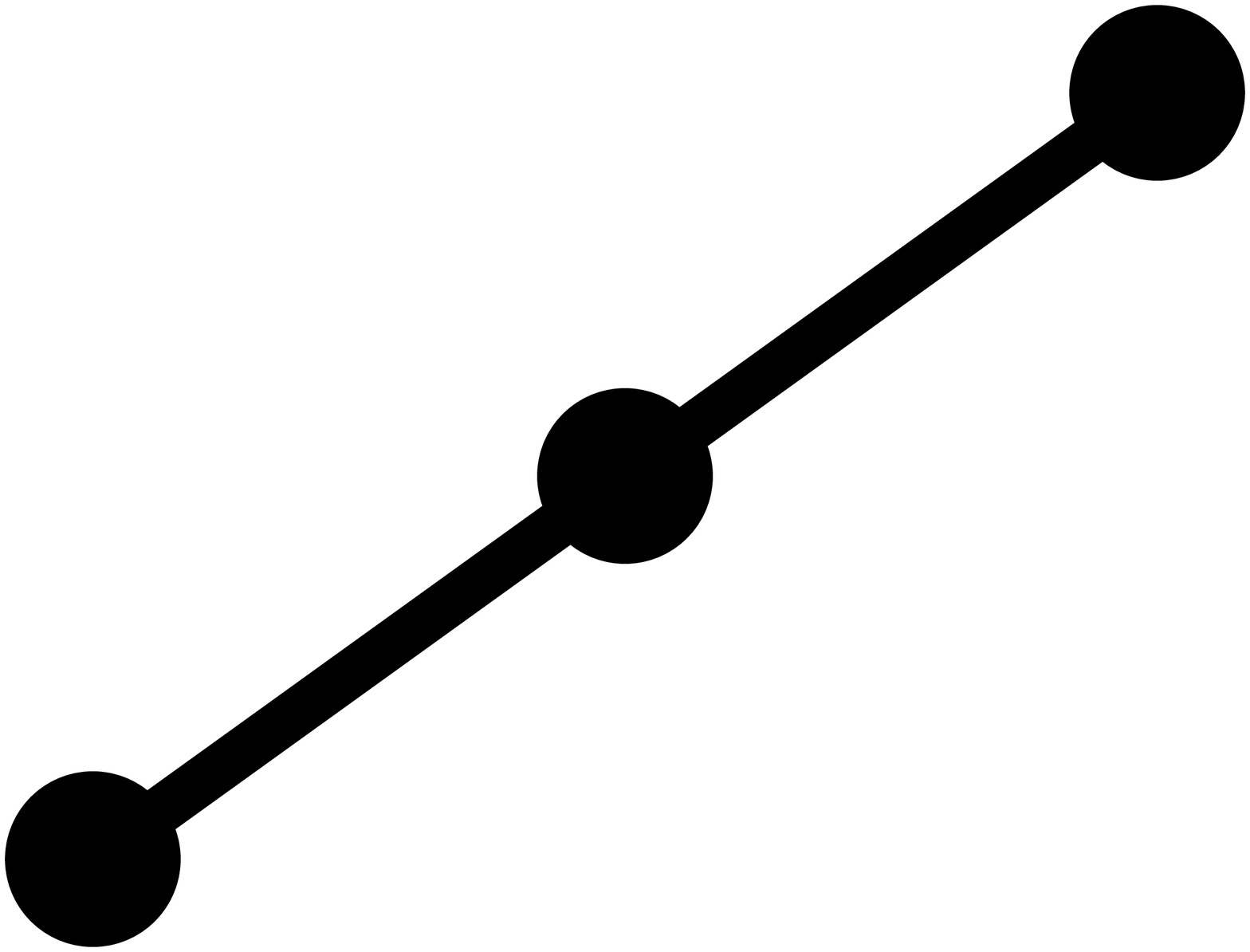}}} & MCM & \cellcolor[gray]{0.9}{0.010} & 0.018 & 0.041 & 0.069 & \cellcolor[gray]{0.9}{0.012} & 0.015 & 0.102 & 0.129 & \cellcolor[gray]{0.9}{0.014} & 0.005 & 0.398 & 0.417 \\
 &  & MMCM & \cellcolor[gray]{0.9}{0.004} & 0.055 & 0.000 & 0.058 & \cellcolor[gray]{0.9}{0.011} & 0.070 & 0.000 & 0.081 & \cellcolor[gray]{0.9}{0.088} & 0.109 & 0.000 & 0.197 \\
 &  & KW & \multicolumn{1}{c}{--} & \multicolumn{1}{c}{--} & \multicolumn{1}{c}{--} & 0.058 & \multicolumn{1}{c}{--} & \multicolumn{1}{c}{--} & \multicolumn{1}{c}{--} & 0.132 & \multicolumn{1}{c}{--} & \multicolumn{1}{c}{--} & \multicolumn{1}{c}{--} & 0.484 \\ \hhline{~--------------}
 & \multirow{3}{*}{\shortstack{(ii) dominant\\ \includegraphics[width=.5cm]{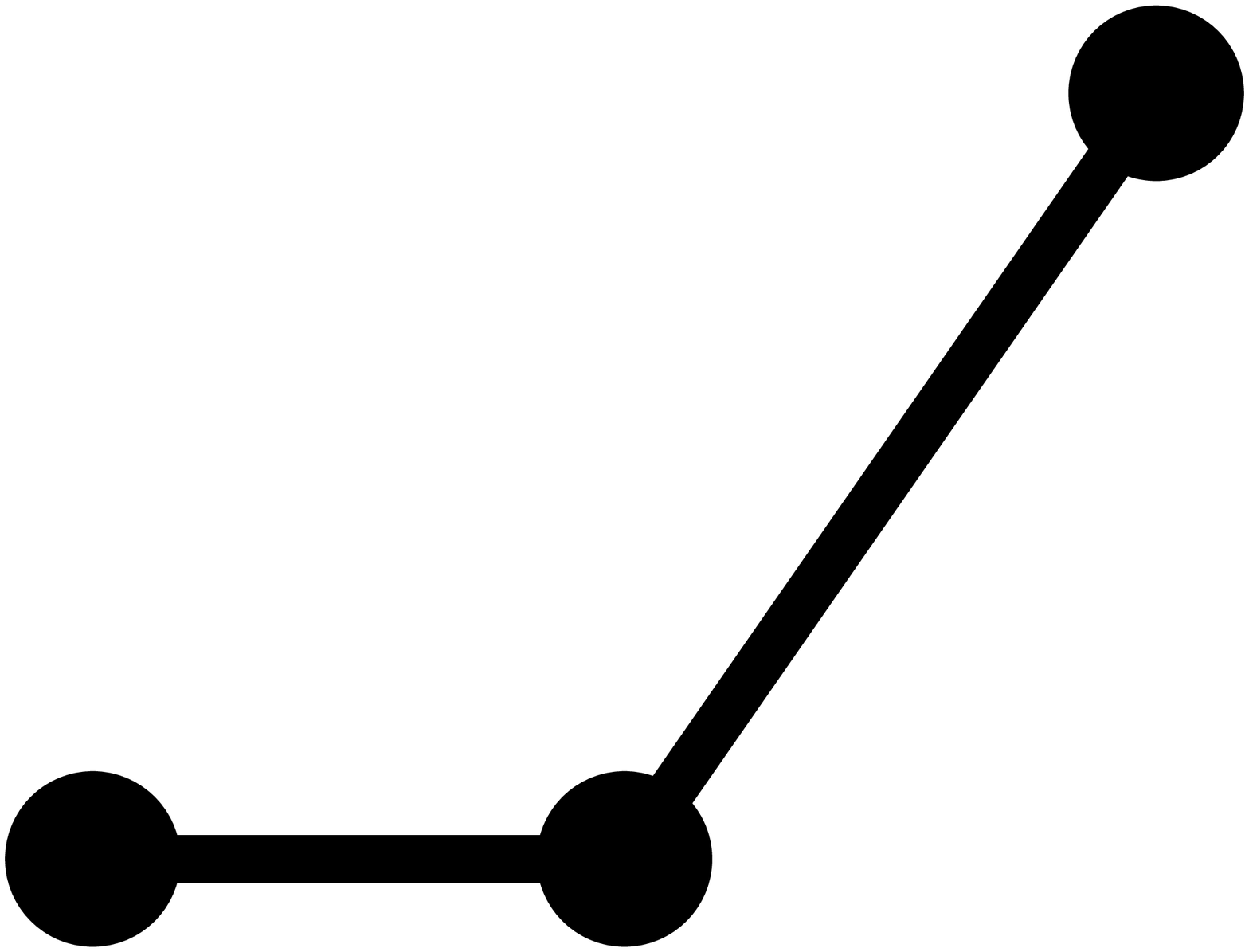}}} & MCM & 0.011 & \cellcolor[gray]{0.9}{0.026} & 0.027 & 0.064 & 0.018 & \cellcolor[gray]{0.9}{0.045} & 0.039 & 0.103 & 0.058 & \cellcolor[gray]{0.9}{0.120} & 0.091 & 0.270 \\
 &  & MMCM & 0.002 & \cellcolor[gray]{0.9}{0.060} & 0.000 & 0.062 & 0.002 & \cellcolor[gray]{0.9}{0.104} & 0.000 & 0.106 & 0.004 & \cellcolor[gray]{0.9}{0.271} & 0.000 & 0.275 \\
 &  & KW & \multicolumn{1}{c}{--} & \multicolumn{1}{c}{--} & \multicolumn{1}{c}{--} & 0.040 & \multicolumn{1}{c}{--} & \multicolumn{1}{c}{--} & \multicolumn{1}{c}{--} & 0.053 & \multicolumn{1}{c}{--} & \multicolumn{1}{c}{--} & \multicolumn{1}{c}{--} & 0.105 \\ \hhline{~--------------}
 & \multirow{3}{*}{\shortstack{(iii) recessive\\ \includegraphics[width=.5cm]{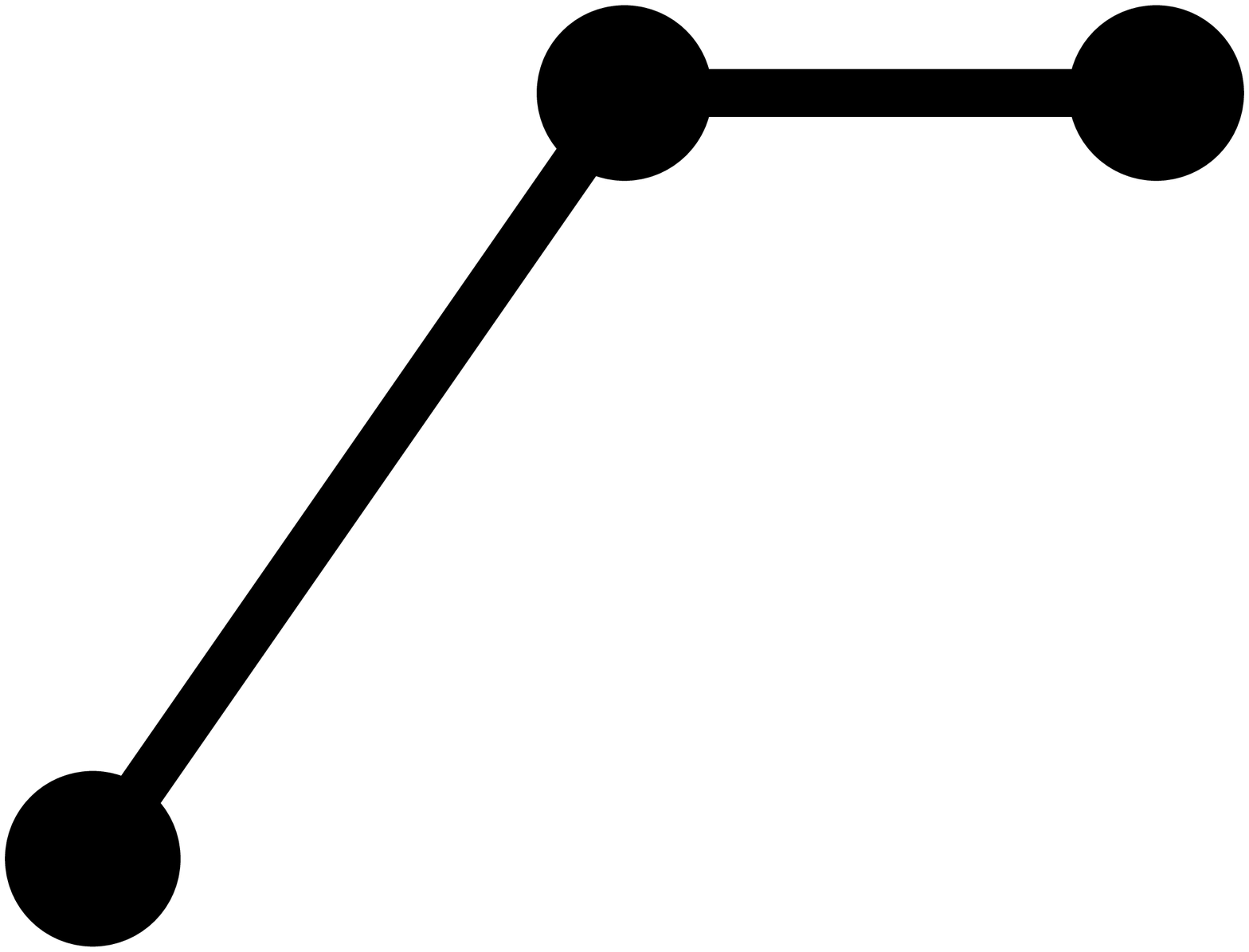}}} & MCM & 0.007 & 0.014 & \cellcolor[gray]{0.9}{0.062} & 0.083 & 0.005 & 0.008 & \cellcolor[gray]{0.9}{0.196} & 0.208 & 0.000 & 0.002 & \cellcolor[gray]{0.9}{0.664} & 0.666 \\
 &  & MMCM & 0.005 & 0.049 & \cellcolor[gray]{0.9}{0.000} & 0.054 & 0.029 & 0.044 & \cellcolor[gray]{0.9}{0.000} & 0.074 & 0.161 & 0.011 & \cellcolor[gray]{0.9}{0.031} & 0.203 \\
 &  & KW & \multicolumn{1}{c}{--} & \multicolumn{1}{c}{--} & \multicolumn{1}{c}{--} & 0.116 & \multicolumn{1}{c}{--} & \multicolumn{1}{c}{--} & \multicolumn{1}{c}{--} & 0.399 & \multicolumn{1}{c}{--} & \multicolumn{1}{c}{--} & \multicolumn{1}{c}{--} & 0.954 \\ \hhline{---------------}
\multirow{9}{*}{0.25} & \multirow{3}{*}{\shortstack{(i) additive\\ \includegraphics[width=.5cm]{type-a.eps}}} & MCM & \cellcolor[gray]{0.9}{0.021} & 0.020 & 0.062 & 0.104 & \cellcolor[gray]{0.9}{0.046} & 0.021 & 0.211 & 0.278 & \cellcolor[gray]{0.9}{0.066} & 0.008 & 0.727 & 0.801 \\
 &  & MMCM & \cellcolor[gray]{0.9}{0.020} & 0.062 & 0.001 & 0.083 & \cellcolor[gray]{0.9}{0.083} & 0.092 & 0.006 & 0.181 & \cellcolor[gray]{0.9}{0.438} & 0.115 & 0.054 & 0.607 \\
 &  & KW & \multicolumn{1}{c}{--} & \multicolumn{1}{c}{--} & \multicolumn{1}{c}{--} & 0.090 & \multicolumn{1}{c}{--} & \multicolumn{1}{c}{--} & \multicolumn{1}{c}{--} & 0.243 & \multicolumn{1}{c}{--} & \multicolumn{1}{c}{--} & \multicolumn{1}{c}{--} & 0.760 \\ \hhline{~--------------}
 & \multirow{3}{*}{\shortstack{(ii) dominant\\ \includegraphics[width=.5cm]{type-b.eps}}} & MCM & 0.020 & \cellcolor[gray]{0.9}{0.038} & 0.032 & 0.089 & 0.061 & \cellcolor[gray]{0.9}{0.114} & 0.053 & 0.228 & 0.205 & \cellcolor[gray]{0.9}{0.390} & 0.083 & 0.678 \\
 &  & MMCM & 0.010 & \cellcolor[gray]{0.9}{0.083} & 0.000 & 0.093 & 0.022 & \cellcolor[gray]{0.9}{0.224} & 0.000 & 0.246 & 0.045 & \cellcolor[gray]{0.9}{0.668} & 0.000 & 0.713 \\
 &  & KW & \multicolumn{1}{c}{--} & \multicolumn{1}{c}{--} & \multicolumn{1}{c}{--} & 0.075 & \multicolumn{1}{c}{--} & \multicolumn{1}{c}{--} & \multicolumn{1}{c}{--} & 0.168 & \multicolumn{1}{c}{--} & \multicolumn{1}{c}{--} & \multicolumn{1}{c}{--} & 0.550 \\ \hhline{~--------------}
 & \multirow{3}{*}{\shortstack{(iii) recessive\\ \includegraphics[width=.5cm]{type-c.eps}}} & MCM & 0.014 & 0.012 & \cellcolor[gray]{0.9}{0.110} & 0.136 & 0.010 & 0.004 & \cellcolor[gray]{0.9}{0.425} & 0.438 & 0.000 & 0.000 & \cellcolor[gray]{0.9}{0.969} & 0.970 \\
 &  & MMCM & 0.033 & 0.040 & \cellcolor[gray]{0.9}{0.003} & 0.075 & 0.116 & 0.022 & \cellcolor[gray]{0.9}{0.050} & 0.188 & 0.196 & 0.001 & \cellcolor[gray]{0.9}{0.616} & 0.812 \\
 &  & KW & \multicolumn{1}{c}{--} & \multicolumn{1}{c}{--} & \multicolumn{1}{c}{--} & 0.171 & \multicolumn{1}{c}{--} & \multicolumn{1}{c}{--} & \multicolumn{1}{c}{--} & 0.556 & \multicolumn{1}{c}{--} & \multicolumn{1}{c}{--} & \multicolumn{1}{c}{--} & 0.994 \\ \hhline{---------------}
\multirow{9}{*}{0.33} & \multirow{3}{*}{\shortstack{(i) additive\\ \includegraphics[width=.5cm]{type-a.eps}}} & MCM & \cellcolor[gray]{0.9}{0.030} & 0.026 & 0.060 & 0.116 & \cellcolor[gray]{0.9}{0.089} & 0.037 & 0.215 & 0.342 & \cellcolor[gray]{0.9}{0.199} & 0.020 & 0.672 & 0.891 \\
 &  & MMCM & \cellcolor[gray]{0.9}{0.038} & 0.058 & 0.005 & 0.102 & \cellcolor[gray]{0.9}{0.149} & 0.103 & 0.028 & 0.279 & \cellcolor[gray]{0.9}{0.609} & 0.106 & 0.110 & 0.824 \\
 &  & KW & \multicolumn{1}{c}{--} & \multicolumn{1}{c}{--} & \multicolumn{1}{c}{--} & 0.099 & \multicolumn{1}{c}{--} & \multicolumn{1}{c}{--} & \multicolumn{1}{c}{--} & 0.274 & \multicolumn{1}{c}{--} & \multicolumn{1}{c}{--} & \multicolumn{1}{c}{--} & 0.831 \\ \hhline{~--------------}
 & \multirow{3}{*}{\shortstack{(ii) dominant\\ \includegraphics[width=.5cm]{type-b.eps}}} & MCM & 0.030 & \cellcolor[gray]{0.9}{0.054} & 0.029 & 0.113 & 0.087 & \cellcolor[gray]{0.9}{0.192} & 0.041 & 0.320 & 0.223 & \cellcolor[gray]{0.9}{0.610} & 0.029 & 0.862 \\
 &  & MMCM & 0.025 & \cellcolor[gray]{0.9}{0.095} & 0.002 & 0.122 & 0.053 & \cellcolor[gray]{0.9}{0.302} & 0.001 & 0.357 & 0.072 & \cellcolor[gray]{0.9}{0.818} & 0.000 & 0.890 \\
 &  & KW & \multicolumn{1}{c}{--} & \multicolumn{1}{c}{--} & \multicolumn{1}{c}{--} & 0.093 & \multicolumn{1}{c}{--} & \multicolumn{1}{c}{--} & \multicolumn{1}{c}{--} & 0.255 & \multicolumn{1}{c}{--} & \multicolumn{1}{c}{--} & \multicolumn{1}{c}{--} & 0.791 \\ \hhline{~--------------}
 & \multirow{3}{*}{\shortstack{(iii) recessive\\ \includegraphics[width=.5cm]{type-c.eps}}} & MCM & 0.020 & 0.014 & \cellcolor[gray]{0.9}{0.124} & 0.159 & 0.022 & 0.004 & \cellcolor[gray]{0.9}{0.489} & 0.515 & 0.001 & 0.000 & \cellcolor[gray]{0.9}{0.987} & 0.988 \\
 &  & MMCM & 0.050 & 0.033 & \cellcolor[gray]{0.9}{0.020} & 0.103 & 0.145 & 0.014 & \cellcolor[gray]{0.9}{0.172} & 0.332 & 0.151 & 0.001 & \cellcolor[gray]{0.9}{0.804} & 0.955 \\
 &  & KW & \multicolumn{1}{c}{--} & \multicolumn{1}{c}{--} & \multicolumn{1}{c}{--} & 0.172 & \multicolumn{1}{c}{--} & \multicolumn{1}{c}{--} & \multicolumn{1}{c}{--} & 0.556 & \multicolumn{1}{c}{--} & \multicolumn{1}{c}{--} & \multicolumn{1}{c}{--} & 0.993 \\ \hhline{---------------}
\multirow{9}{*}{0.50} & \multirow{3}{*}{\shortstack{(i) additive\\ \includegraphics[width=.5cm]{type-a.eps}}} & MCM & \cellcolor[gray]{0.9}{0.041} & 0.045 & 0.045 & 0.132 & \cellcolor[gray]{0.9}{0.155} & 0.120 & 0.117 & 0.392 & \cellcolor[gray]{0.9}{0.529} & 0.195 & 0.199 & 0.923 \\
 &  & MMCM & \cellcolor[gray]{0.9}{0.062} & 0.035 & 0.035 & 0.132 & \cellcolor[gray]{0.9}{0.226} & 0.088 & 0.084 & 0.398 & \cellcolor[gray]{0.9}{0.692} & 0.116 & 0.119 & 0.927 \\
 &  & KW & \multicolumn{1}{c}{--} & \multicolumn{1}{c}{--} & \multicolumn{1}{c}{--} & 0.104 & \multicolumn{1}{c}{--} & \multicolumn{1}{c}{--} & \multicolumn{1}{c}{--} & 0.304 & \multicolumn{1}{c}{--} & \multicolumn{1}{c}{--} & \multicolumn{1}{c}{--} & 0.864 \\ \hhline{~--------------}
 & \multirow{3}{*}{\shortstack{(ii) dominant\\ \includegraphics[width=.5cm]{type-b.eps}}} & MCM & 0.034 & \cellcolor[gray]{0.9}{0.095} & 0.020 & 0.149 & 0.079 & \cellcolor[gray]{0.9}{0.392} & 0.013 & 0.483 & 0.049 & \cellcolor[gray]{0.9}{0.928} & 0.000 & 0.977 \\
 &  & MMCM & 0.053 & \cellcolor[gray]{0.9}{0.080} & 0.015 & 0.148 & 0.127 & \cellcolor[gray]{0.9}{0.346} & 0.007 & 0.480 & 0.101 & \cellcolor[gray]{0.9}{0.875} & 0.000 & 0.977 \\
 &  & KW & \multicolumn{1}{c}{--} & \multicolumn{1}{c}{--} & \multicolumn{1}{c}{--} & 0.137 & \multicolumn{1}{c}{--} & \multicolumn{1}{c}{--} & \multicolumn{1}{c}{--} & 0.443 & \multicolumn{1}{c}{--} & \multicolumn{1}{c}{--} & \multicolumn{1}{c}{--} & 0.968 \\ \hhline{~--------------}
 & \multirow{3}{*}{\shortstack{(iii) recessive\\ \includegraphics[width=.5cm]{type-c.eps}}} & MCM & 0.035 & 0.019 & \cellcolor[gray]{0.9}{0.102} & 0.156 & 0.077 & 0.014 & \cellcolor[gray]{0.9}{0.388} & 0.480 & 0.043 & 0.000 & \cellcolor[gray]{0.9}{0.935} & 0.978 \\
 &  & MMCM & 0.056 & 0.013 & \cellcolor[gray]{0.9}{0.087} & 0.156 & 0.125 & 0.008 & \cellcolor[gray]{0.9}{0.345} & 0.478 & 0.098 & 0.000 & \cellcolor[gray]{0.9}{0.879} & 0.977 \\
 &  & KW & \multicolumn{1}{c}{--} & \multicolumn{1}{c}{--} & \multicolumn{1}{c}{--} & 0.144 & \multicolumn{1}{c}{--} & \multicolumn{1}{c}{--} & \multicolumn{1}{c}{--} & 0.439 & \multicolumn{1}{c}{--} & \multicolumn{1}{c}{--} & \multicolumn{1}{c}{--} & 0.967 \\ \hhline{---------------}
\multicolumn{15}{l}{\scriptsize \shortstack[l]{
\vspace{-1.7pt}
\\ \vspace{-3.6pt}
Abbreviations: MAF, minor allele frequency; MCM, maximum contrast method; MMCM, modified maximum contrast method; KW, Kruskal--Wallis
\\ \vspace{-3.6pt}
test. Shaded region shows positive predictive value ($\hat{R}_{\mathrm{TP}}$) for detection of true response patterns.}}
\end{tabular}
\end{center}
\vspace{-17pt}
\end{table} %

\end{landscape}

\begin{landscape}
\begin{table}[!p]
\vspace{-10pt}
\begin{center}
\caption{\small The $\hat{R}_{\mathrm{P}}$ (power) and $\hat{R}_{\mathrm{TP}}$ (positive predicted value) for various response patterns $(n = 300)$}
\label{tab:SimResult300}
\scriptsize
\begin{tabular}{cccrrrrrrrrrrrr} \hline
 &  &  & \multicolumn{4}{c}{$\Delta$ = 0.25} & \multicolumn{4}{c}{$\Delta$ = 0.50} & \multicolumn{4}{c}{$\Delta$ = 1.00} \\ \hhline{---------------}
MAF & True situation & Method & \multicolumn{1}{c}{(i)} & \multicolumn{1}{c}{(ii)} & \multicolumn{1}{c}{(iii)} & \multicolumn{1}{c}{$\hat{R}_{\mathrm{P}}$} & \multicolumn{1}{c}{(i)} & \multicolumn{1}{c}{(ii)} & \multicolumn{1}{c}{(iii)} & \multicolumn{1}{c}{$\hat{R}_{\mathrm{P}}$} & \multicolumn{1}{c}{(i)} & \multicolumn{1}{c}{(ii)} & \multicolumn{1}{c}{(iii)} & \multicolumn{1}{c}{$\hat{R}_{\mathrm{P}}$} \\ \hhline{---------------}
\multirow{9}{*}{0.12} & \multirow{3}{*}{\shortstack{(i) additive\\ \includegraphics[width=.5cm]{type-a.eps}}} & MCM & \cellcolor[gray]{0.9}{0.010} & 0.015 & 0.073 & 0.098 & \cellcolor[gray]{0.9}{0.013} & 0.007 & 0.273 & 0.293 & \cellcolor[gray]{0.9}{0.001} & 0.000 & 0.834 & 0.835 \\
 &  & MMCM & \cellcolor[gray]{0.9}{0.005} & 0.063 & 0.000 & 0.068 & \cellcolor[gray]{0.9}{0.043} & 0.100 & 0.000 & 0.143 & \cellcolor[gray]{0.9}{0.381} & 0.094 & 0.005 & 0.480 \\
 &  & KW & \multicolumn{1}{c}{--} & \multicolumn{1}{c}{--} & \multicolumn{1}{c}{--} & 0.121 & \multicolumn{1}{c}{--} & \multicolumn{1}{c}{--} & \multicolumn{1}{c}{--} & 0.392 & \multicolumn{1}{c}{--} & \multicolumn{1}{c}{--} & \multicolumn{1}{c}{--} & 0.942 \\ \hhline{~--------------}
 & \multirow{3}{*}{\shortstack{(ii) dominant\\ \includegraphics[width=.5cm]{type-b.eps}}} & MCM & 0.015 & \cellcolor[gray]{0.9}{0.035} & 0.033 & 0.083 & 0.036 & \cellcolor[gray]{0.9}{0.081} & 0.070 & 0.188 & 0.135 & \cellcolor[gray]{0.9}{0.273} & 0.179 & 0.587 \\
 &  & MMCM & 0.001 & \cellcolor[gray]{0.9}{0.083} & 0.000 & 0.084 & 0.002 & \cellcolor[gray]{0.9}{0.190} & 0.000 & 0.192 & 0.006 & \cellcolor[gray]{0.9}{0.590} & 0.000 & 0.596 \\
 &  & KW & \multicolumn{1}{c}{--} & \multicolumn{1}{c}{--} & \multicolumn{1}{c}{--} & 0.064 & \multicolumn{1}{c}{--} & \multicolumn{1}{c}{--} & \multicolumn{1}{c}{--} & 0.131 & \multicolumn{1}{c}{--} & \multicolumn{1}{c}{--} & \multicolumn{1}{c}{--} & 0.424 \\ \hhline{~--------------}
 & \multirow{3}{*}{\shortstack{(iii) recessive\\ \includegraphics[width=.5cm]{type-c.eps}}} & MCM & 0.004 & 0.010 & \cellcolor[gray]{0.9}{0.133} & 0.147 & 0.000 & 0.003 & \cellcolor[gray]{0.9}{0.477} & 0.481 & 0.000 & 0.001 & \cellcolor[gray]{0.9}{0.977} & 0.977 \\
 &  & MMCM & 0.015 & 0.047 & \cellcolor[gray]{0.9}{0.000} & 0.062 & 0.109 & 0.020 & \cellcolor[gray]{0.9}{0.004} & 0.134 & 0.220 & 0.001 & \cellcolor[gray]{0.9}{0.454} & 0.676 \\
 &  & KW & \multicolumn{1}{c}{--} & \multicolumn{1}{c}{--} & \multicolumn{1}{c}{--} & 0.321 & \multicolumn{1}{c}{--} & \multicolumn{1}{c}{--} & \multicolumn{1}{c}{--} & 0.887 & \multicolumn{1}{c}{--} & \multicolumn{1}{c}{--} & \multicolumn{1}{c}{--} & 1.000 \\ \hhline{---------------}
\multirow{9}{*}{0.25} & \multirow{3}{*}{\shortstack{(i) additive\\ \includegraphics[width=.5cm]{type-a.eps}}} & MCM & \cellcolor[gray]{0.9}{0.035} & 0.021 & 0.154 & 0.210 & \cellcolor[gray]{0.9}{0.059} & 0.011 & 0.589 & 0.659 & \cellcolor[gray]{0.9}{0.008} & 0.000 & 0.990 & 0.998 \\
 &  & MMCM & \cellcolor[gray]{0.9}{0.055} & 0.085 & 0.002 & 0.141 & \cellcolor[gray]{0.9}{0.306} & 0.124 & 0.024 & 0.454 & \cellcolor[gray]{0.9}{0.850} & 0.031 & 0.100 & 0.981 \\
 &  & KW & \multicolumn{1}{c}{--} & \multicolumn{1}{c}{--} & \multicolumn{1}{c}{--} & 0.193 & \multicolumn{1}{c}{--} & \multicolumn{1}{c}{--} & \multicolumn{1}{c}{--} & 0.629 & \multicolumn{1}{c}{--} & \multicolumn{1}{c}{--} & \multicolumn{1}{c}{--} & 0.997 \\ \hhline{~--------------}
 & \multirow{3}{*}{\shortstack{(ii) dominant\\ \includegraphics[width=.5cm]{type-b.eps}}} & MCM & 0.041 & \cellcolor[gray]{0.9}{0.080} & 0.046 & 0.167 & 0.150 & \cellcolor[gray]{0.9}{0.289} & 0.085 & 0.523 & 0.326 & \cellcolor[gray]{0.9}{0.613} & 0.043 & 0.982 \\
 &  & MMCM & 0.013 & \cellcolor[gray]{0.9}{0.166} & 0.000 & 0.179 & 0.037 & \cellcolor[gray]{0.9}{0.516} & 0.000 & 0.553 & 0.014 & \cellcolor[gray]{0.9}{0.971} & 0.000 & 0.986 \\
 &  & KW & \multicolumn{1}{c}{--} & \multicolumn{1}{c}{--} & \multicolumn{1}{c}{--} & 0.129 & \multicolumn{1}{c}{--} & \multicolumn{1}{c}{--} & \multicolumn{1}{c}{--} & 0.419 & \multicolumn{1}{c}{--} & \multicolumn{1}{c}{--} & \multicolumn{1}{c}{--} & 0.958 \\ \hhline{~--------------}
 & \multirow{3}{*}{\shortstack{(iii) recessive\\ \includegraphics[width=.5cm]{type-c.eps}}} & MCM & 0.010 & 0.005 & \cellcolor[gray]{0.9}{0.316} & 0.331 & 0.000 & 0.001 & \cellcolor[gray]{0.9}{0.892} & 0.893 & 0.000 & 0.000 & \cellcolor[gray]{0.9}{1.000} & 1.000 \\
 &  & MMCM & 0.085 & 0.026 & \cellcolor[gray]{0.9}{0.018} & 0.129 & 0.211 & 0.003 & \cellcolor[gray]{0.9}{0.386} & 0.600 & 0.078 & 0.000 & \cellcolor[gray]{0.9}{0.922} & 1.000 \\
 &  & KW & \multicolumn{1}{c}{--} & \multicolumn{1}{c}{--} & \multicolumn{1}{c}{--} & 0.450 & \multicolumn{1}{c}{--} & \multicolumn{1}{c}{--} & \multicolumn{1}{c}{--} & 0.971 & \multicolumn{1}{c}{--} & \multicolumn{1}{c}{--} & \multicolumn{1}{c}{--} & 1.000 \\ \hhline{---------------}
\multirow{9}{*}{0.33} & \multirow{3}{*}{\shortstack{(i) additive\\ \includegraphics[width=.5cm]{type-a.eps}}} & MCM & \cellcolor[gray]{0.9}{0.071} & 0.034 & 0.169 & 0.275 & \cellcolor[gray]{0.9}{0.173} & 0.027 & 0.583 & 0.783 & \cellcolor[gray]{0.9}{0.096} & 0.000 & 0.904 & 1.000 \\
 &  & MMCM & \cellcolor[gray]{0.9}{0.108} & 0.094 & 0.018 & 0.220 & \cellcolor[gray]{0.9}{0.478} & 0.121 & 0.087 & 0.686 & \cellcolor[gray]{0.9}{0.909} & 0.023 & 0.067 & 0.999 \\
 &  & KW & \multicolumn{1}{c}{--} & \multicolumn{1}{c}{--} & \multicolumn{1}{c}{--} & 0.230 & \multicolumn{1}{c}{--} & \multicolumn{1}{c}{--} & \multicolumn{1}{c}{--} & 0.712 & \multicolumn{1}{c}{--} & \multicolumn{1}{c}{--} & \multicolumn{1}{c}{--} & 1.000 \\ \hhline{~--------------}
 & \multirow{3}{*}{\shortstack{(ii) dominant\\ \includegraphics[width=.5cm]{type-b.eps}}} & MCM & 0.066 & \cellcolor[gray]{0.9}{0.136} & 0.042 & 0.244 & 0.207 & \cellcolor[gray]{0.9}{0.489} & 0.040 & 0.736 & 0.188 & \cellcolor[gray]{0.9}{0.810} & 0.002 & 1.000 \\
 &  & MMCM & 0.043 & \cellcolor[gray]{0.9}{0.226} & 0.001 & 0.271 & 0.074 & \cellcolor[gray]{0.9}{0.703} & 0.001 & 0.777 & 0.011 & \cellcolor[gray]{0.9}{0.989} & 0.000 & 1.000 \\
 &  & KW & \multicolumn{1}{c}{--} & \multicolumn{1}{c}{--} & \multicolumn{1}{c}{--} & 0.201 & \multicolumn{1}{c}{--} & \multicolumn{1}{c}{--} & \multicolumn{1}{c}{--} & 0.658 & \multicolumn{1}{c}{--} & \multicolumn{1}{c}{--} & \multicolumn{1}{c}{--} & 0.999 \\ \hhline{~--------------}
 & \multirow{3}{*}{\shortstack{(iii) recessive\\ \includegraphics[width=.5cm]{type-c.eps}}} & MCM & 0.022 & 0.005 & \cellcolor[gray]{0.9}{0.379} & 0.406 & 0.003 & 0.000 & \cellcolor[gray]{0.9}{0.946} & 0.949 & 0.000 & 0.000 & \cellcolor[gray]{0.9}{1.000} & 1.000 \\
 &  & MMCM & 0.121 & 0.019 & \cellcolor[gray]{0.9}{0.099} & 0.239 & 0.182 & 0.001 & \cellcolor[gray]{0.9}{0.670} & 0.853 & 0.038 & 0.000 & \cellcolor[gray]{0.9}{0.962} & 1.000 \\
 &  & KW & \multicolumn{1}{c}{--} & \multicolumn{1}{c}{--} & \multicolumn{1}{c}{--} & 0.449 & \multicolumn{1}{c}{--} & \multicolumn{1}{c}{--} & \multicolumn{1}{c}{--} & 0.969 & \multicolumn{1}{c}{--} & \multicolumn{1}{c}{--} & \multicolumn{1}{c}{--} & 1.000 \\ \hhline{---------------}
\multirow{9}{*}{0.50} & \multirow{3}{*}{\shortstack{(i) additive\\ \includegraphics[width=.5cm]{type-a.eps}}} & MCM & \cellcolor[gray]{0.9}{0.115} & 0.096 & 0.102 & 0.312 & \cellcolor[gray]{0.9}{0.437} & 0.200 & 0.202 & 0.839 & \cellcolor[gray]{0.9}{0.817} & 0.090 & 0.093 & 1.000 \\
 &  & MMCM & \cellcolor[gray]{0.9}{0.169} & 0.071 & 0.077 & 0.317 & \cellcolor[gray]{0.9}{0.589} & 0.127 & 0.129 & 0.845 & \cellcolor[gray]{0.9}{0.945} & 0.028 & 0.028 & 1.000 \\
 &  & KW & \multicolumn{1}{c}{--} & \multicolumn{1}{c}{--} & \multicolumn{1}{c}{--} & 0.244 & \multicolumn{1}{c}{--} & \multicolumn{1}{c}{--} & \multicolumn{1}{c}{--} & 0.761 & \multicolumn{1}{c}{--} & \multicolumn{1}{c}{--} & \multicolumn{1}{c}{--} & 1.000 \\ \hhline{~--------------}
 & \multirow{3}{*}{\shortstack{(ii) dominant\\ \includegraphics[width=.5cm]{type-b.eps}}} & MCM & 0.072 & \cellcolor[gray]{0.9}{0.290} & 0.016 & 0.379 & 0.065 & \cellcolor[gray]{0.9}{0.865} & 0.002 & 0.932 & 0.002 & \cellcolor[gray]{0.9}{0.998} & 0.000 & 1.000 \\
 &  & MMCM & 0.112 & \cellcolor[gray]{0.9}{0.255} & 0.010 & 0.377 & 0.123 & \cellcolor[gray]{0.9}{0.806} & 0.001 & 0.930 & 0.016 & \cellcolor[gray]{0.9}{0.985} & 0.000 & 1.000 \\
 &  & KW & \multicolumn{1}{c}{--} & \multicolumn{1}{c}{--} & \multicolumn{1}{c}{--} & 0.351 & \multicolumn{1}{c}{--} & \multicolumn{1}{c}{--} & \multicolumn{1}{c}{--} & 0.915 & \multicolumn{1}{c}{--} & \multicolumn{1}{c}{--} & \multicolumn{1}{c}{--} & 1.000 \\ \hhline{~--------------}
 & \multirow{3}{*}{\shortstack{(iii) recessive\\ \includegraphics[width=.5cm]{type-c.eps}}} & MCM & 0.068 & 0.014 & \cellcolor[gray]{0.9}{0.300} & 0.382 & 0.064 & 0.002 & \cellcolor[gray]{0.9}{0.867} & 0.933 & 0.002 & 0.000 & \cellcolor[gray]{0.9}{0.998} & 1.000 \\
 &  & MMCM & 0.106 & 0.009 & \cellcolor[gray]{0.9}{0.264} & 0.379 & 0.124 & 0.001 & \cellcolor[gray]{0.9}{0.807} & 0.931 & 0.013 & 0.000 & \cellcolor[gray]{0.9}{0.987} & 1.000 \\
 &  & KW & \multicolumn{1}{c}{--} & \multicolumn{1}{c}{--} & \multicolumn{1}{c}{--} & 0.354 & \multicolumn{1}{c}{--} & \multicolumn{1}{c}{--} & \multicolumn{1}{c}{--} & 0.917 & \multicolumn{1}{c}{--} & \multicolumn{1}{c}{--} & \multicolumn{1}{c}{--} & 1.000 \\ \hhline{---------------}
\multicolumn{15}{l}{\scriptsize \shortstack[l]{
\vspace{-1.7pt}
\\ \vspace{-3.6pt}
Abbreviations: MAF, minor allele frequency; MCM, maximum contrast method; MMCM, modified maximum contrast method; KW, Kruskal--Wallis
\\ \vspace{-3.6pt}
test. Shaded region shows positive predictive value ($\hat{R}_{\mathrm{TP}}$) for detection of true response patterns.}}
\end{tabular}
\end{center}
\vspace{-17pt}
\end{table} %

\end{landscape}

\begin{table}[htb]
\begin{center}
\caption{False positive rates for valley response pattern $(n = 100)$}
\label{tab:FP100}
\small
\begin{tabular}{cccrrr} \hline
MAF & Response pattern &  & \multicolumn{1}{c}{$\Delta = 0.25$} & \multicolumn{1}{c}{$\Delta = 0.50$} & \multicolumn{1}{c}{$\Delta = 1.00$} \\ \hline
\multirow{3}{*}{0.12} & \multirow{3}{*}{\shortstack{(iv) valley\\ \includegraphics[width=.5cm]{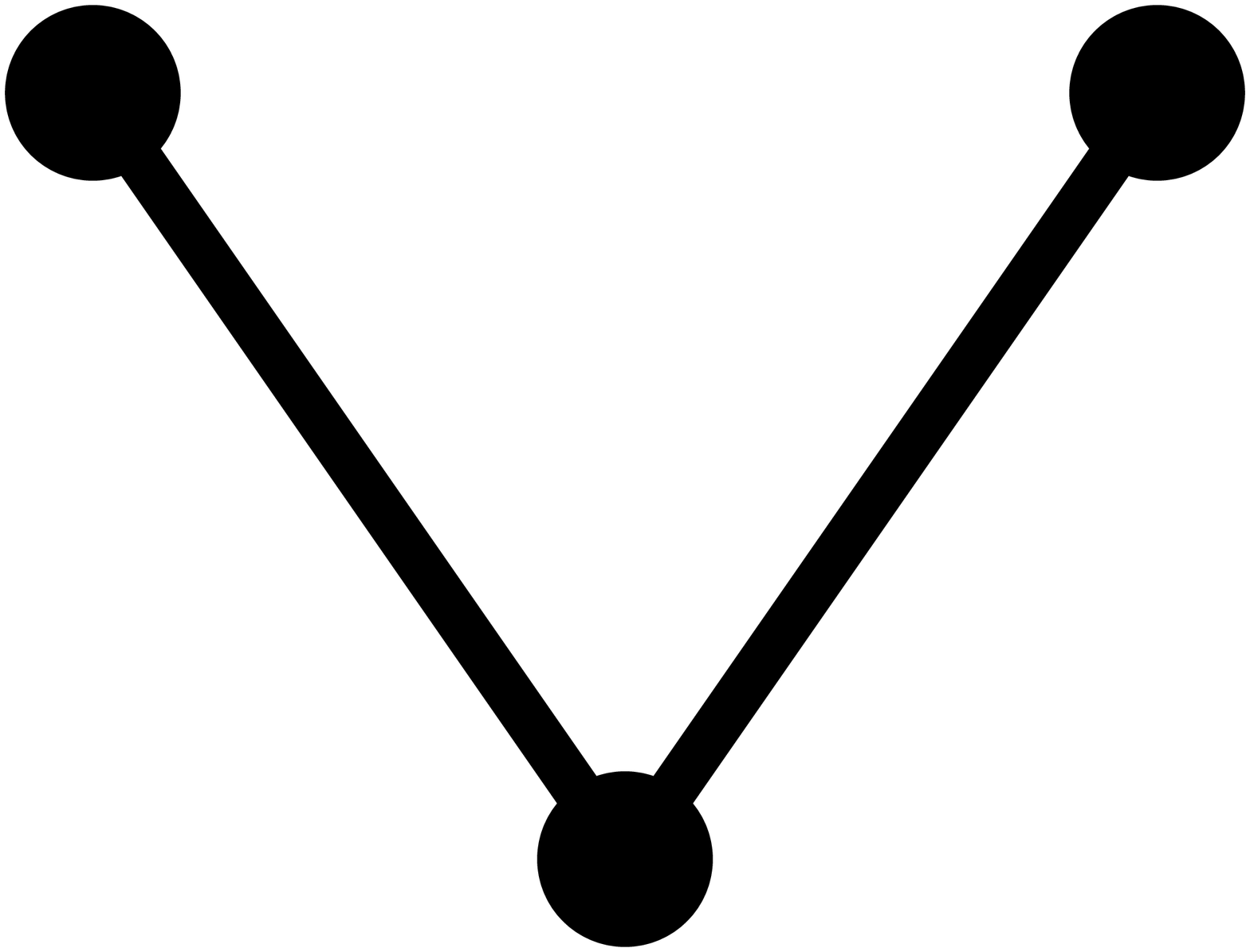}}} & MCM & 0.066 & 0.103 & 0.273 \\
 &  & MMCM & \cellcolor[gray]{0.9}{0.055} & \cellcolor[gray]{0.9}{0.064} & \cellcolor[gray]{0.9}{0.119} \\
 &  & KW & 0.118 & 0.380 & 0.936 \\ \hline
\multirow{3}{*}{0.25} & \multirow{3}{*}{\shortstack{(iv) valley\\ \includegraphics[width=.5cm]{type-d.eps}}} & MCM & 0.082 & 0.183 & 0.604 \\
 &  & MMCM & \cellcolor[gray]{0.9}{0.062} & \cellcolor[gray]{0.9}{0.111} & \cellcolor[gray]{0.9}{0.363} \\
 &  & KW & 0.161 & 0.537 & 0.989 \\ \hline
\multirow{3}{*}{0.33} & \multirow{3}{*}{\shortstack{(iv) valley\\ \includegraphics[width=.5cm]{type-d.eps}}} & MCM & 0.091 & 0.223 & 0.742 \\
 &  & MMCM & \cellcolor[gray]{0.9}{0.076} & \cellcolor[gray]{0.9}{0.164} & \cellcolor[gray]{0.9}{0.608} \\
 &  & KW & 0.170 & 0.560 & 0.993 \\ \hline
\multirow{3}{*}{0.50} & \multirow{3}{*}{\shortstack{(iv) valley\\ \includegraphics[width=.5cm]{type-d.eps}}} & MCM & 0.095 & 0.248 & 0.804 \\
 &  & MMCM & \cellcolor[gray]{0.9}{0.091} & \cellcolor[gray]{0.9}{0.238} & \cellcolor[gray]{0.9}{0.792} \\
 &  & KW & 0.173 & 0.559 & 0.994 \\ \hline
\multicolumn{6}{l}{\scriptsize \shortstack[l]{\\Abbreviations: MAF, minor allele frequency; MCM, maximum contrast method; MMCM, \\modified maximum contrast method; KW, Kruskal--Wallis test. Shaded region is minimum \\false positive rate at each MAF and $\Delta$.}}
\end{tabular}
\end{center}
\end{table} %

\begin{table}[htb]
\begin{center}
\caption{False positive rates for the valley response pattern $(n = 300)$}
\label{tab:FP300}
\small
\begin{tabular}{cccrrr} \hline
MAF & Response pattern &  & \multicolumn{1}{c}{$\Delta = 0.25$} & \multicolumn{1}{c}{$\Delta = 0.50$} & \multicolumn{1}{c}{$\Delta = 1.00$} \\ \hline
\multirow{3}{*}{0.12} & \multirow{3}{*}{\shortstack{(iv) valley\\ \includegraphics[width=.5cm]{type-d.eps}}} & MCM & 0.091 & 0.198 & 0.629 \\
 &  & MMCM & \cellcolor[gray]{0.9}{0.062} & \cellcolor[gray]{0.9}{0.091} & \cellcolor[gray]{0.9}{0.245} \\
 &  & KW & 0.306 & 0.869 & 1.000 \\ \hline
\multirow{3}{*}{0.25} & \multirow{3}{*}{\shortstack{(iv) valley\\ \includegraphics[width=.5cm]{type-d.eps}}} & MCM & 0.147 & 0.455 & 0.996 \\
 &  & MMCM & \cellcolor[gray]{0.9}{0.091} & \cellcolor[gray]{0.9}{0.247} & \cellcolor[gray]{0.9}{0.944} \\
 &  & KW & 0.425 & 0.964 & 1.000 \\ \hline
\multirow{3}{*}{0.33} & \multirow{3}{*}{\shortstack{(iv) valley\\ \includegraphics[width=.5cm]{type-d.eps}}} & MCM & 0.175 & 0.584 & 1.000 \\
 &  & MMCM & \cellcolor[gray]{0.9}{0.128} & \cellcolor[gray]{0.9}{0.439} & \cellcolor[gray]{0.9}{0.999} \\
 &  & KW & 0.448 & 0.968 & 1.000 \\ \hline
\multirow{3}{*}{0.50} & \multirow{3}{*}{\shortstack{(iv) valley\\ \includegraphics[width=.5cm]{type-d.eps}}} & MCM & 0.197 & 0.664 & 1.000 \\
 &  & MMCM & \cellcolor[gray]{0.9}{0.190} & \cellcolor[gray]{0.9}{0.648} & 1.000 \\
 &  & KW & 0.452 & 0.971 & 1.000 \\ \hline
\multicolumn{6}{l}{\scriptsize \shortstack[l]{\\Abbreviations: MAF, minor allele frequency; MCM, maximum contrast method; MMCM, \\modified maximum contrast method; KW, Kruskal--Wallis test. Shaded region is minimum \\false positive rate at each MAF and $\Delta$.}}
\end{tabular}
\end{center}
\end{table} %

\end{document}